\documentclass[aps,prb,twocolumn,citeautoscript,groupedaddress]{revtex4-1}

\usepackage[colorlinks,urlcolor=blue,citecolor=blue]{hyperref}
\usepackage{color}
\usepackage{graphicx}
\usepackage{amsmath}
\usepackage{amssymb}
\usepackage{mathtools}

\usepackage[dvipsnames]{xcolor}

\begin{document}

\title{Pre-equilibrium stopping and charge capture in proton-irradiated aluminum sheets}

\author{Alina Kononov}
\affiliation{Department of Physics, University of Illinois at Urbana-Champaign, Urbana, IL 61801, USA}
\author{Andr\'{e} Schleife}
\email{schleife@illinois.edu}
\affiliation{Department of Materials Science and Engineering, University of Illinois at Urbana-Champaign, Urbana, IL 61801, USA}
\affiliation{Materials Research Laboratory, University of Illinois at Urbana-Champaign, Urbana, IL 61801, USA}
\affiliation{National Center for Supercomputing Applications, University of Illinois at Urbana-Champaign, Urbana, IL 61801, USA}

\date{\today}

\begin{abstract}
We present a first-principles study of pre-equilibrium stopping power and projectile charge capture in thin aluminum sheets irradiated by 6\,\---\,60\,keV protons.
Our time-dependent density functional theory calculations reveal enhanced stopping power compared to bulk aluminum, particularly near the entrance layers.
We propose the additional excitation channel of surface plasma oscillations as the most plausible explanation for this behavior.
We also introduce a novel technique to compute the orbital-resolved charge state of a proton projectile after transmission through the sheet.
Our results provide insight into the dynamics of orbital occupations after the projectile exits the aluminum sheet and have important implications for advancing radiation hardness and focused-ion beam techniques, especially for few-layer materials.
\end{abstract}

\maketitle

\section{Introduction}

Although ion irradiation represents a fundamental means of manipulating materials and probing their properties\cite{Notte:2007,Utke:2008,Ritter:2013,Hlawacek:2014,Fox:2015,Li:2017:2D}, a detailed theoretical understanding of the interaction between an energetic charged particle and induced electronic excitations in a solid has proven challenging.
The Bethe formula\cite{Bethe:1930,Sigmund:2008} models stopping power $S$=$-dE/dz$ (i.e., the energy deposited per penetration depth) for fast projectiles, while the Fermi-Teller model\cite{Fermi:1947} applies to slow projectiles.
The Lindhard stopping formula\cite{Lindhard:1964} from linear-response theory and its extensions\cite{Wang:1998} accurately account for the small density perturbations produced by fast, low-$Z$ projectiles.
However, these existing analytical models\cite{Bethe:1930,Sigmund:2008,Fermi:1947,Lindhard:1964,Wang:1998} rely on ambiguous parameters such as projectile charge $Z$ or participating electron density $n$ which not only have multiple meaningful definitions, but may also evolve dynamically during the projectile's transit \cite{Correa:2018}.

In addition, upon entering the material, quantities such as stopping power and the projectile's charge state are not initially identical to the steady-state bulk values presumably achieved as the projectile moves through a thick target \cite{Lee:2020}.
This leads to pre-equilibrium effects that cannot be ignored when understanding electronic stopping near surfaces or in thin target materials of only a few atomic layers.
In these, the projectile may not reach an equilibrium charge state at all, since it may need to traverse many layers before doing so \cite{Lee:2020}.
While pre-equilibrium effects should occur even for light ions, they should be most prominent for highly charged projectiles with a large difference between initial and equilibrated charge state.

Indeed, several experiments on highly charged ions impacting carbon-based materials with thicknesses of only a few nm or less reported that the projectile's initial charge influences energy and charge transfer distributions \cite{Hattass:1999,Wilhelm:2014} and even material damage characteristics \cite{Schenkel:1997,Ritter:2013}.
Similarly, the response of aluminum foils to ion irradiation has been shown to depend on both ion charge and foil thickness \cite{Sun:1996,Brandt:1973}.
Such studies inferred projectile charge equilibration time scales smaller than 10\,fs and length scales shorter than 10\,nm \cite{Brandt:1973,Schenkel:1997,Hattass:1999,Wilhelm:2017,Wilhelm:2018}.
Sensitivity of electron emission to incident ion charge was shown even for $\sim$\,100\,nm thick carbon foils and attributed to pre-equilibrium stopping and projectile charge \cite{Koschar:1989}.

These experimental observations of pre-equilibrium effects inspired exponential decay models for projectile charge equilibration \cite{Brandt:1973,Hattass:1999}.
Since experiments cannot access the projectile's charge state within the material, studies evaluating such models typically compare their predictions to measurements of the projectile's charge after transmission through the sample \cite{Hattass:1999,Wilhelm:2018}.
However, this exit charge state may not be equivalent to the projectile's charge state inside the material.
Overall, the transition and equilibration of the ion into the bulk regime is still poorly understood and requires further study.

Conversely, for bulk materials under ion irradiation, stopping power has been fairly well-studied experimentally \cite{Mertens:1982,Semrad:1986,Moller:2004}.
In addition, modern first-principles simulations have provided a detailed description of the energy and charge dynamics in bulk, as evident from many recent studies of electronic stopping power in metals\cite{Correa:2012,Schleife:2015,Quashie:2016,Caro:2017,Ullah:2018}, semiconductors\cite{Lim:2016,Yost:2016,Yost:2017,Lee:2018,Kang:2019,Lee:2019,Lee:2020}, and insulators\cite{Pruneda:2007,Mao:2014:LiF,Li:2017} which were enabled by the rise of high performance computing.
However, it remains difficult for experiments to access the poorly understood pre-equilibrium behavior of stopping power near a surface or for materials thinner than the length scale of projectile charge equilibration.

To improve our understanding of pre-equilibrium behavior, computational modeling of thin materials under ion irradiation is a promising alternative, as it offers greater spatial and temporal resolution than currently achievable experimentally.
However, modeling an ion's interaction with a surface requires the inclusion of a sufficiently large vacuum region and sufficiently large material surface, greatly increasing computational cost.
In addition, extracting observables from the simulated time-dependent electron density to describe the charge dynamics instigated by ion-induced electronic excitations presents a further challenge.
Since detailed understanding of pre-equilibrium behavior is currently absent, we aim to study these effects in the present work.

To this end, we used a first-principles computational approach to calculate and analyze electronic stopping and projectile charge state as protons traverse aluminum sheets.
We focus on protons with kinetic energies of 6\,--\,60\,keV that move along a channeling trajectory through 0.8\,--\,2.4\,nm  of aluminum, corresponding to 4\,--\,12 atomic layers.
The wealth of existing literature on the electronic response of bulk aluminum to proton irradiation makes this an ideal system to study dynamical behavior near ion-irradiated surfaces.
In particular, the stopping power of protons in bulk aluminum has been studied extensively both experimentally\cite{Mertens:1982,Semrad:1986,Moller:2004} and theoretically\cite{Penalba:1992,Quijada:2007,Correa:2012,Zeb:2013,Schleife:2015}.
This existing wisdom provides opportunities both to validate our results for the bulk limit and to clearly identify surface effects and pre-equilibrium behavior.
Moreover, light-ion irradiation is particularly well-suited for first-principles studies because they experience relatively little scattering in the host material, resulting in long, straight trajectories \cite{Notte:2007}.
Using a proton projectile further allows us to accurately calculate the charge captured by the projectile using analytic hydrogen orbitals, as we discuss below.

The remainder of this paper is structured as follows:
In Section \ref{sec:methods} we outline the first-principles computational approach used here, and in Section \ref{sec:density_analysis} we describe the post-processing methods developed and used to extract charge capture from time-dependent electron density data.
Section \ref{sec:results} presents the results obtained for pre-equilibrium stopping power and charge capture, and Section \ref{sec:conclusion} summarizes the conclusions drawn from this study.

\section{Computational Methods \label{sec:methods}}

We used real-time time-dependent density functional theory (TDDFT) \cite{Runge:1984,Marques:2004,Marques:2006,Ullrich:2011,Ullrich:2014} as implemented \cite{Schleife:2012,Schleife:2014,Schleife:2015} in the Qbox/Qb@ll code \cite{Gygi:2008,qball:2017} to simulate the real-time non-adiabatic electronic-ion dynamics as a proton traverses thin aluminum sheets.
Different sheet thicknesses consisting of 4\,--\,12 atomic layers were studied, where one layer corresponds to about 0.2\,nm.
The Kohn-Sham (KS) orbitals were expanded in a plane-wave basis with a cutoff energy of 50\,Ry.
The electron-ion interaction was described using a Troullier-Martins pseudopotential \cite{Troullier:1991} with 11 valence electrons for aluminum and a Hamann-Schl{\"u}ter-Chiang-Vanderbilt pseudopotential \cite{Vanderbilt:1985} for the proton projectile.
The large simulation cells used in this work, typically $38\times 38\times 150$\,a$_0^3$, allow reciprocal-space sampling using the $\Gamma$ point only.
The adiabatic local density approximation \cite{Zangwill:1980,Zangwill:1981} was used to describe exchange and correlation.
The electronic ground state of the aluminum sheet served as the initial condition of the time-dependent calculations, and it was computed using density functional theory with 100 empty states and a Fermi electron temperature of 100\,K in order to accelerate self-consistent convergence of the metallic ground state.

\begin{figure*}
\begin{center}
\includegraphics[width=0.8\textwidth]{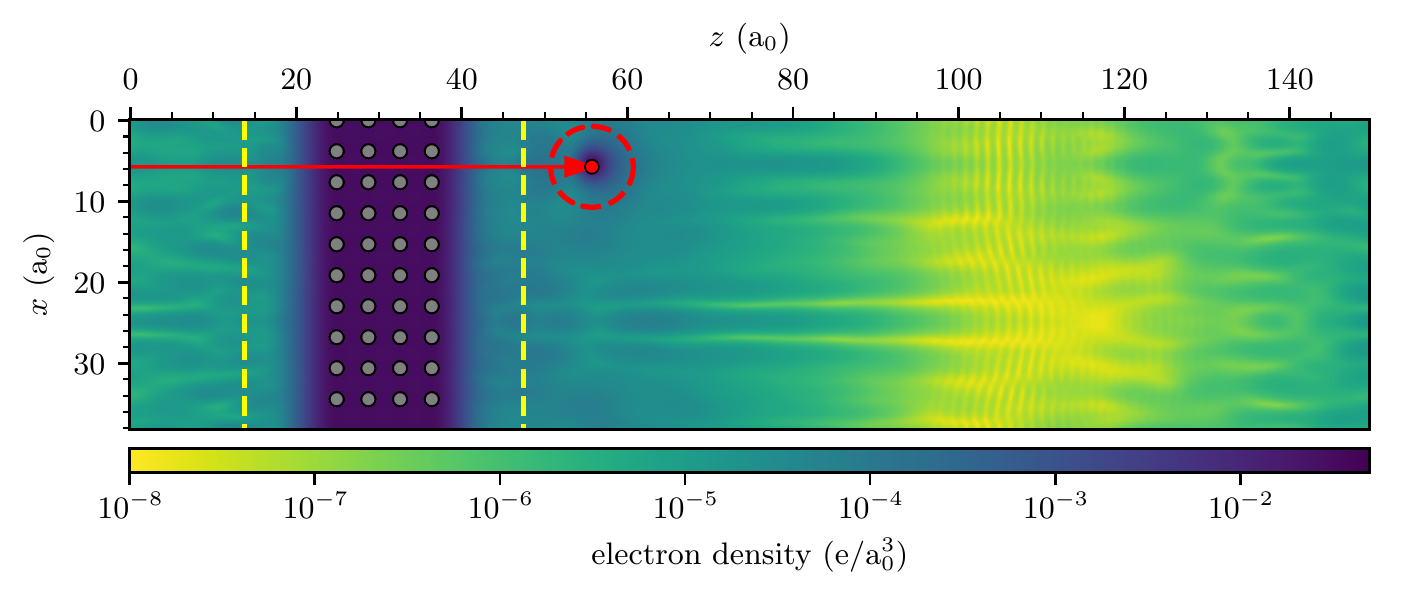}
\end{center}
\caption{
\label{fig:eden}(Color online.)
Illustration of our simulation cell at a time of 0.75\,fs after a proton with a velocity of 1\,at.\,u.\ impacts a 4-layer thick aluminum sheet.
A slice of the electron density is overlaid with projected positions of aluminum atoms in gray and the projectile along with its trajectory in red.
Electron density has been emitted into the vacuum and captured by the projectile.
Dashed yellow lines indicate the aluminum-vacuum boundaries used for calculating the sheet's dipole moment, and the 5\,a$_0$ projectile radius\cite{Zhao:2015} used to compare our charge capture method with volume partitioning is indicated by the red dashed circle (see Section \ref{sec:density_analysis}).
}
\end{figure*}

Due to the few-fs time scale of our time-dependent simulations, ions do not have enough time to respond to the forces they experience.
Thus, we fix the positions of the aluminum ions and maintain a constant velocity of the proton projectile.
The proton starts at a distance of at least 25\,a$_0$ from the aluminum sheet and traverses it along a [100] channeling trajectory (see Fig.\ \ref{fig:eden}).
The enforced time reversal symmetry (ETRS) integrator\cite{Castro:2004,Draeger:2017} with a time step of 0.042\,at.\,u.\ (1.04\,as) was used to propagate the Kohn-Sham orbitals, a choice shown to produce high numerical accuracy for similar systems\cite{Kang:2019}.
The cutoff energy, treatment of semi-core electrons, target atom dynamics, and channeling projectile trajectory used in this work were shown in Ref.\ \onlinecite{Schleife:2015} to produce accurate stopping power results for proton-irradiated bulk aluminum within the velocity regime presently considered.
Accuracy and convergence of the time propagation for these large simulation cells is addressed in the Supplemental Material of this paper (see Fig.\ \ref{fig:integrators}).
From our real-time TDDFT simulations we obtain time-dependent electron densities which we analyze further using the approaches discussed in the following section.
Inputs and outputs from our simulations are available at the Materials Data Facility \cite{MDF,data}.

\section{\label{sec:density_analysis}Electron Density Analysis}

It is a central goal of this work to quantify and analyze the charge state of the projectile both inside and outside the aluminum target material.
Information about the projectile's charge state could be estimated by analyzing coefficients of the KS orbitals in an atomic orbital basis\cite{Mulliken:1955,Knospe:1999,Knospe:2000,Pruneda:2007,Isborn:2007}, by projecting the KS orbitals onto atomic orbitals\cite{Miyamoto:2008}, or by considering probabilities that the KS orbitals are localized within a given volume of interest \cite{Reinhard:1999,Ullrich:2000,Wang:2011,Wang:2012,Gao:2013}.
However, many of these methods have well-known shortcomings, most notably that they rely directly on the single-particle, non-interacting KS orbitals which have no rigorous physical meaning.

Instead, a method that provides the projectile charge state as a functional of only the total electron density---a real, physical quantity which, in principle, determines all observables\cite{Runge:1984}---would be preferable.
Existing methods that extract atomic charges directly from the electron density typically rely on volume\cite{Bader:1994} or charge\cite{Manz:2016, Gabaldon_limas:2016} partitioning, which either assume a definite boundary between the captured electrons and nearby free electrons or neglect free electrons altogether.
In this context, we apply the DDEC6\cite{Manz:2016,Gabaldon_limas:2016} charge partitioning method in this work to calculate the effective projectile charge state as the proton traverses the aluminum sheet.
However, this technique is not applicable in the presence of free electrons and, hence, we find that it overestimates the number of electrons bound to the projectile once it emerges from the material into the vacuum containing emitted electrons (see Fig.\ \ref{fig:DDEC_vs_fit} of the Supplemental Material).

Since the electron distribution emitted into vacuum, not including those captured by the projectile, should not be assigned to any atoms, a different method is required to calculate charge captured by the projectile after traversing the material.
Once the projectile has left the target material, a common strategy simply integrates the electron density within a sphere centered at the projectile \cite{Reinhard:1998, Miyamoto:2008:general, Avendano:2012, Mao:2014:Gr, Ojanpera:2014, Zhao:2015, Li:2017}.
However, the radius chosen for this sphere (for instance, 5\,a$_0$ in Ref.\ \onlinecite{Zhao:2015}) defines an artificial, discrete boundary between electrons captured by the projectile and free electrons in the vacuum which, depending on the radius chosen and the occupied projectile orbitals, could either falsely include emitted electrons or falsely exclude higher energy captured electrons.

\begin{figure}
\includegraphics[width=\columnwidth]{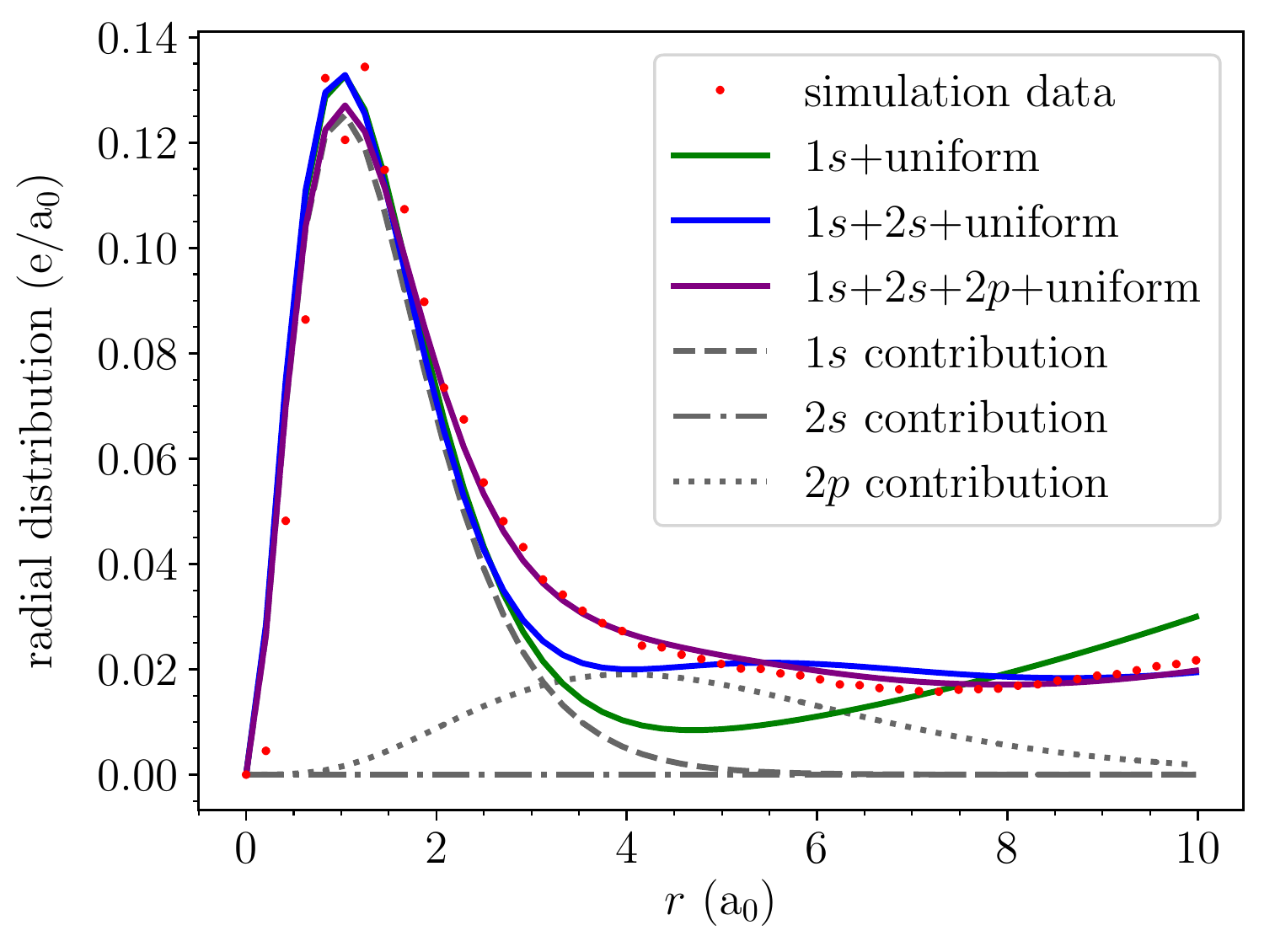}
\caption{
\label{fig:orb_fit}(Color online.)
Curve fitting used to determine the orbital occupations of the proton projectile and to calculate the charge it captured.
Red circles show simulation results for the radial distribution of the electron density around the projectile (1\,at.\,u.\ of velocity) 1.2\,fs after it exits from a 4-layer aluminum sheet.
Least squares fits to this data using different analytic hydrogen orbitals and a uniform background charge are shown in green, blue, and purple.
Radial distributions for analytic 1$s$, 2$s$, and 2$p$ hydrogen orbitals are shown in gray; they are scaled by their respective occupations as obtained from the 1$s$+2$s$+2$p$+uniform fit, which describes the simulation results very well.
}
\end{figure}

Instead, we present a novel, physically motivated method of calculating the charge captured by a proton and also the orbital distribution of the captured charge as a functional of the electron density.
We first obtain the radial distribution $n(r,t)$ of the electron density around the projectile, in units of e/a$_0$, by integrating the electron density $n(\mathbf{r},t)$ over spherical shells $S$,
\begin{equation}
\label{eq:nr}
n(r,t)=\frac{1}{\Delta r}\int_{S(r,\mathbf{R}(t))} n(\mathbf{r},t) \; dr^3.
\end{equation}
Here, $\Delta r$ is the smallest real-space grid spacing in each simulation and $S(r,\mathbf{R}(t))$ is the spherical shell of thickness $\Delta r$ and radius $r$, centered at the projectile's position $\mathbf{R}(t)$.
Integrating $n(r,t)$ from Eq.\ \eqref{eq:nr} again over $r$ would give the number of electrons within a sphere around the projectile.
Using the electron densities from our simulations, represented on a real-space grid, we evaluate the following discrete version of Eq.\ \eqref{eq:nr} to compute $n(r,t)$:
\begin{equation}
n(r_i,t) = \frac{1}{\Delta r} \sum_{\mathclap{r_{i-1}< |\mathbf{r}-\mathbf{R}(t)| \leq r_i }} n(\mathbf{r},t) \; \Delta V
\end{equation}
where $r_i=i\Delta r$ ranges from $0$ to 10\,a$_0$ and $\Delta V$ is the volume of each grid cell.
We then fit the resulting radial distribution to a linear combination of analytic radial distributions of hydrogen orbitals (1$s$, 2$s$, and 2$p$) and a uniform background density to account for nearby free electrons.
The resulting fits capture the numerically calculated radial distributions extremely well (see Fig.\ \ref{fig:orb_fit}), with $R^2$ values generally above 0.9.
In the Supplemental Material, we show that the DFT orbitals of an isolated H$^+$ ion closely match the analytical hydrogen orbitals (see Fig.\ \ref{fig:H_orbitals}) and that including even higher energy orbitals has a negligible effect on the orbital occupations and the total charge transfer (see Fig.\ \ref{fig:moreorbitals}).

Finally, in order to analyze plasma oscillations induced in the aluminum sheet, we compute the out-of-plane dipole moment
\begin{equation}
\label{eq:dm}
d(t) = \int_{V} z \; n(\mathbf{r} ,t) \; dr^3
\end{equation}
where $\hat{z}$ is normal to the aluminum surface and $V$ is the volume occupied by the sheet.
In order to account for the electron density extending from the aluminum surface, we include within $V$ the region within 11\,a$_0$ of the aluminum surface atoms (see yellow dashed lines in \mbox{Fig.\ \ref{fig:eden}}).
Using this cutoff gives less than $6\times 10^{-3}$ electrons in the vacuum at the start of each calculation.

\section{\label{sec:results}Results and Discussion}

\subsection{\label{sec:results_energy}Pre-equilibrium electronic stopping}

A fast projectile impacting a \emph{bulk} target or sufficiently thick foil reaches an equilibrium charge state within a few fs and within the first few nm of material traversed \cite{Hattass:1999,Lee:2020}.
The projectile then experiences equilibrium electronic stopping\cite{Herrmann:1994}, and as a result its velocity decreases over a time scale much longer than the initial charge equilibration.
As the projectile slows down, its equilibrium charge state also evolves as a function of velocity.
However, the situation is more complicated for thin target materials.
Upon approaching and entering the surface, an ion dynamically captures or loses electrons\cite{Bohr:1948}, leading to energy transfer dynamics which have been detected as a dependence of stopping power in thin foils on the initial charge state of the projectile\cite{Ogawa:1993}.
This pre-equilibrium behavior in the projectile charge and stopping power within the material surface would then also influence surface processes such as electron emission.

\begin{figure}
\includegraphics{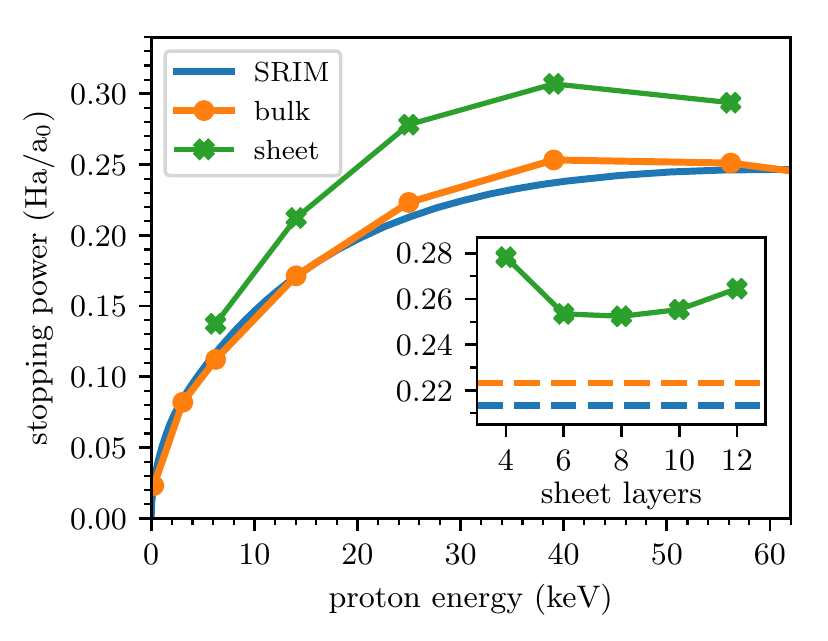}
\caption{
\label{fig:stopping_bulk_SRIM}(Color online.)
Electronic stopping as a function of kinetic energy of a proton projectile in a 4-layer aluminum sheet (green) is higher than in bulk aluminum (TDDFT results \cite{Schleife:2015} in orange and SRIM data \cite{Ziegler:2010} in blue).
Inset: Electronic stopping of a 25\,keV proton (velocity of 1\,at.\,u.) for different aluminum sheet thicknesses.
Dashed lines indicate bulk values from TDDFT \cite{Schleife:2015} (orange) and SRIM \cite{Ziegler:2010} (blue).
For each sheet, average stopping is computed across the two middle layers as the most bulk-like region.
}
\end{figure}

Our results show clear pre-equilibrium effects in the energy transferred from the projectile to the host material from the comparison of the stopping power in thin aluminum sheets to bulk aluminum:
Figure \ref{fig:stopping_bulk_SRIM} shows 13\,--\,25\,\% greater stopping, depending on projectile velocity and sheet thickness, for H$^+$ in the aluminum sheets compared to theoretical \cite{Schleife:2015} and empirically fitted \cite{Ziegler:2010} results for H in bulk aluminum.
Since the stopping powers of H and H$^+$ in bulk aluminum quickly converge toward the same value \cite{Schleife:2015}, this comparison isolates surface effects in the host material.
Furthermore, the inset of Figure \ref{fig:stopping_bulk_SRIM} shows that the stopping power varies with sheet thickness and does not approach the bulk value even for a 12-layer sheet, the thickest one we simulated.

\begin{figure}
\includegraphics[width=\columnwidth]{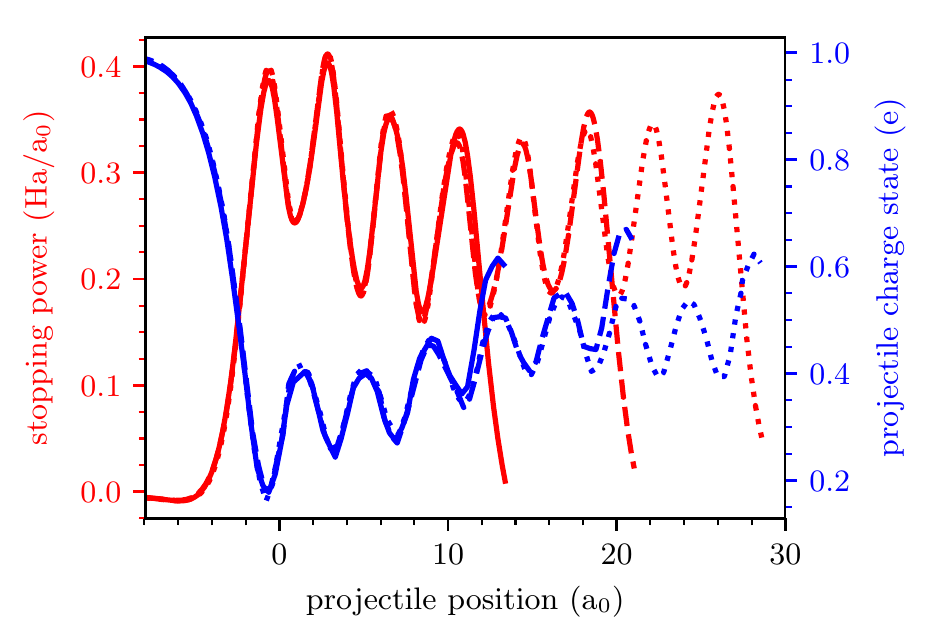}
\caption{
\label{fig:DDEC}(Color online.)
Instantaneous electronic stopping (red) and effective projectile charge state from DDEC6 \cite{Manz:2016,Gabaldon_limas:2016} analysis (blue) for a proton projectile with 1\,at.\,u. of velocity traversing aluminum sheets that are 4, 6, and 8 layers thick (solid, dashed, dotted, respectively).
The entrance layer of aluminum atoms is located 0\,a$_0$.
}
\end{figure}

In order to explain why pre-equilibrium stopping is larger than bulk stopping, we first analyze the projectile charge state dynamics.
This allows us to compare to analytic models that predict that stopping power varies quadratically with projectile charge \cite{Bethe:1930,Sigmund:2008,Fermi:1947,Lindhard:1964,Wang:1998}.
Interestingly, we find from Figure \ref{fig:DDEC} that the instantaneous projectile charge state as calculated by the DDEC6 method\cite{Manz:2016,Gabaldon_limas:2016} is actually anticorrelated with the instantaneous stopping power:
The proton experiences enhanced stopping despite lower effective charge within the first few atomic layers, and the fluctuations arising from lattice periodicity are out of phase, with local maxima in stopping power aligned with local minima in effective charge state and vice versa.
Thus, the dynamical behavior of the projectile charge state appears incompatible with equilibrium stopping power models and does not explain the pre-equilibrium stopping behavior near the material's surface.
However, we also note that charge partitioning schemes may not be capable of accurately resolving the relatively small changes in the charge state occurring here, as the captured electron density is superimposed with the projectile's wake in the host material.

\begin{figure}
\includegraphics[width=\columnwidth]{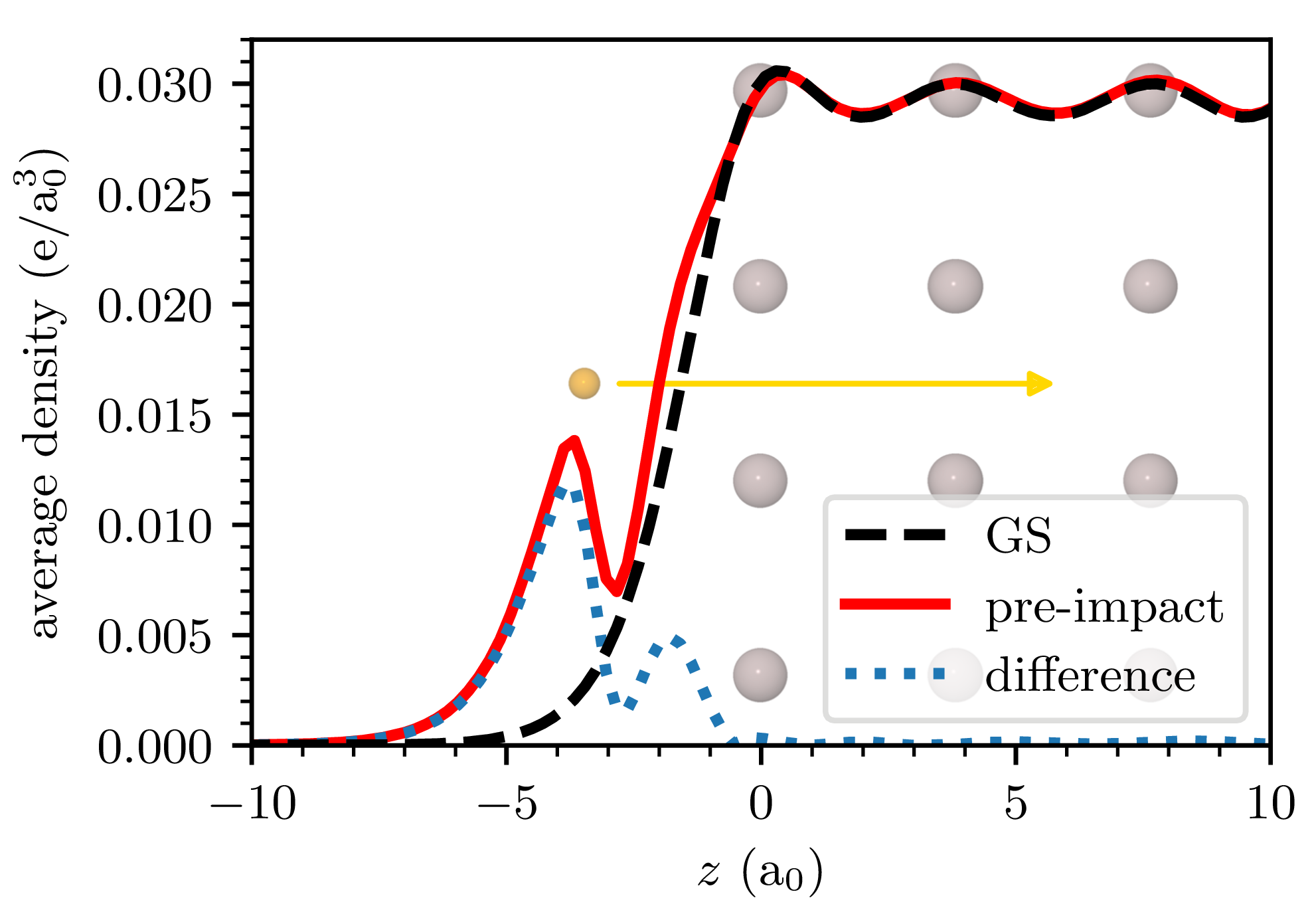}
\caption{
\label{fig:polarization}(Color online.)
Average electron density located within a 1\,a$_0$ radius of the proton's trajectory for the ground-state initial condition (dashed black) and when the proton, approaching with 1\,at.\,u.\ of velocity, is 3.5\,a$_0$ away from impacting a 4-layer aluminum sheet (solid red).
Dotted blue shows the difference between the average densities just before impact and in the ground state.
The positions of the aluminum atoms (proton projectile) are indicated in gray (gold).
}
\end{figure}

Another potential source of enhanced pre-equilibrium stopping, according to analytic models\cite{Bethe:1930,Sigmund:2008,Fermi:1947,Lindhard:1964}, would be higher electron density near the entrance surface.
We indeed observe such higher electron density, arising from the polarization induced on the aluminum sheet during the proton's approach (see Fig.\ \ref{fig:polarization}).
However, we found that before impact, this polarization is highly localized to the very surface, without extending into the first two atomic layers where the enhanced stopping is observed.
Once the projectile enters the material, it again becomes impossible to disentangle its wake from its captured electrons.
Therefore, we cannot definitively attribute the enhanced surface stopping power to surface polarization.

\begin{figure}
\includegraphics{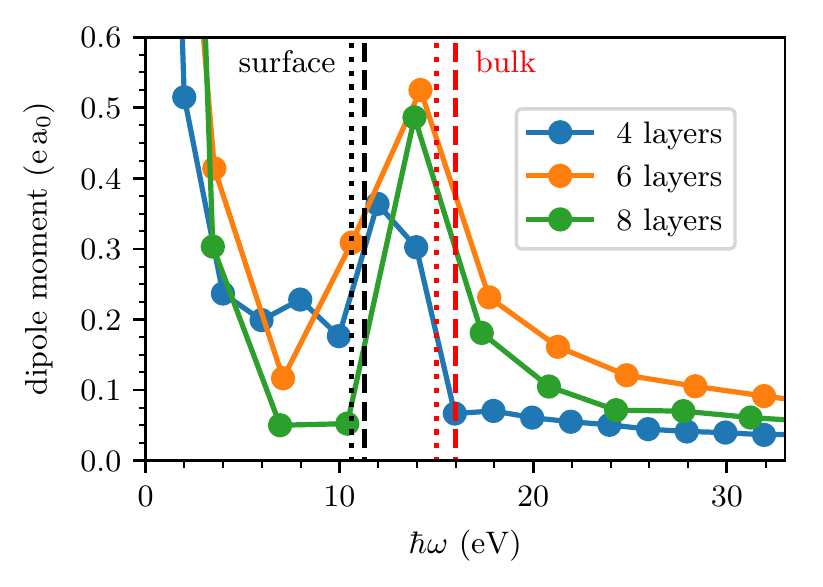}
\caption{
\label{fig:plasmons}(Color online.)
Fourier transform of the time-dependent out-of-plane dipole moment in the aluminum sheets after impact by a proton with 1\,at.\,u.\ of velocity.
Only data at least 0.7\,fs after the projectile exits the material was analyzed in order to isolate plasma oscillations and exclude contributions from emitted electrons.
The dashed (dotted) lines indicate the theoretical (experimental) energies of the bulk and surface plasmons \cite{Pines:1956,Powell:1959,Grosso:2014}.
}
\end{figure}

Finally, the higher stopping may be explained by surface or confinement effects of thin sheets:
Aluminum surfaces have an additional excitation channel in the form of surface plasmons.
In addition, the plasmonic behavior of atomically thin metal films has been shown to deviate from bulk\cite{Maniyara:2019} and predicted to support multiple surface, subsurface, and bulk plasmon modes \cite{Andersen:2012}.
Surface plasmon modes have indeed been predicted to become increasingly important and lead to higher stopping power for incident electrons as film thickness is reduced \cite{Ritchie:1957}.
To investigate this mechanism, we performed a Fourier analysis of the time-dependent out-of-plane dipole moment, Eq.\ \eqref{eq:dm}, and we indeed find indications for plasmon modes located between the bulk and surface plasmon energies (see Fig.\ \ref{fig:plasmons}).
While our frequency resolution is fairly low due to the few-fs time-scale of our simulations, Figure \ref{fig:plasmons} shows that the data for the 4-layer sheet hints at the possibility of two distinct plasmon peaks.
Future studies with significantly longer total propagation time would be needed to unambiguously distinguish specific bulk and surface modes.
Nonetheless, surface plasmon excitations and/or non-bulk plasmonic behavior represent plausible mechanisms for enhanced electronic stopping near an aluminum surface.

We also note that the de Broglie wavelength of electrons at the Fermi surface in aluminum (6.8\,a$_0$) is comparable to the thickness of the aluminum sheets (15\,\---\,92\,a$_0$), suggesting that quantum confinement may affect the thinner sheets studied.
However, since the instantaneous stopping power remains almost identical for sheets with different thicknesses until the projectile reaches the exit surface (see Fig.\ \ref{fig:DDEC}), we conclude that any quantum confinement effects do not significantly influence electronic stopping power in this system.

\subsection{\label{sec:results_charge}Projectile charge capture}

Experimental studies often infer information about pre-equilibrium behavior from charge state distributions of the projectile after transmission through a foil\cite{Burgi:1993,Folkerts:1995,Schenkel:1997,Hattass:1999,Wilhelm:2014}.
Thus, we calculated the number of electrons captured by the projectile after emerging from the aluminum sheets by analyzing the electron density as described in Section \ref{sec:density_analysis}.
Our method also provides information about the sub-fs real-time dynamics of orbital occupations of the captured electrons.

\begin{figure}
\centering
\includegraphics[width=\columnwidth]{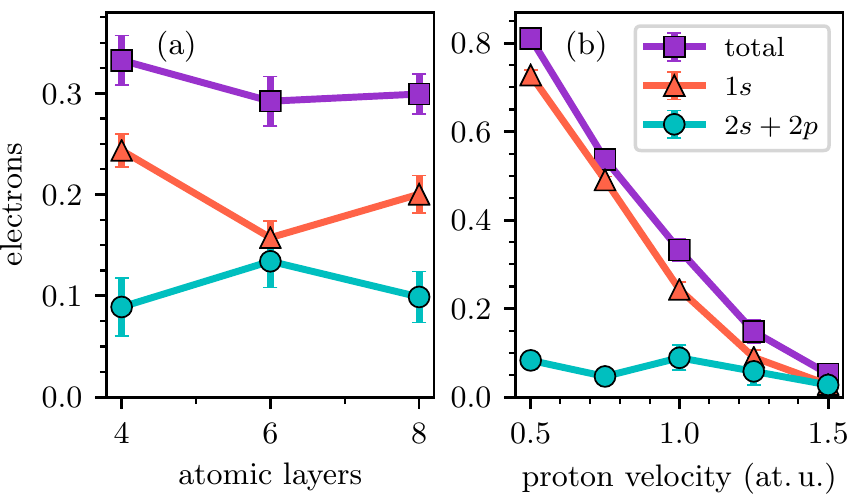}
\caption{
\label{fig:PE_vs_vel_thick}(Color online.)
(a) Total charge captured by the projectile, decomposed into occupations of the $1s$ and ${2s+2p}$ hydrogen orbitals, after a proton with 1\,at.\,u.\ of velocity impacts aluminum sheets of different thickness.
Error bars indicate standard deviations of these time-dependent values. 
(b) The same quantities are shown as a function of projectile velocity for protons impacting 4-layer aluminum.
}
\end{figure}

\begin{figure}
\centering
\includegraphics{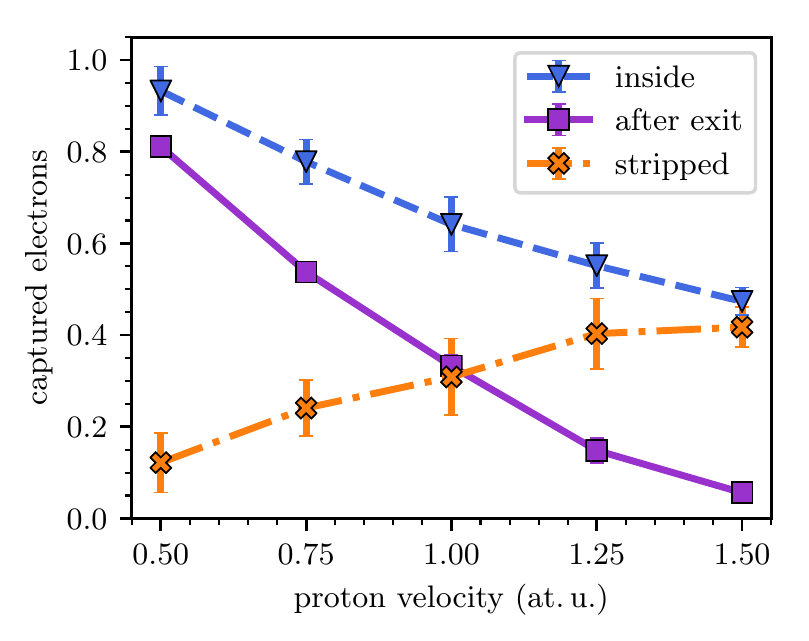}
\caption{
\label{fig:q_stripped}(Color online.)
Average number of electrons captured by the proton projectile inside (blue triangles) and after exiting (purple squares) a 4-layer aluminum sheet.
The difference between these two quantities, the number of electrons stripped at the exit-side surface, is shown in orange exes.
Error bars indicate standard deviations of the time-dependent quantities.
}
\end{figure}

\begin{figure}
\includegraphics{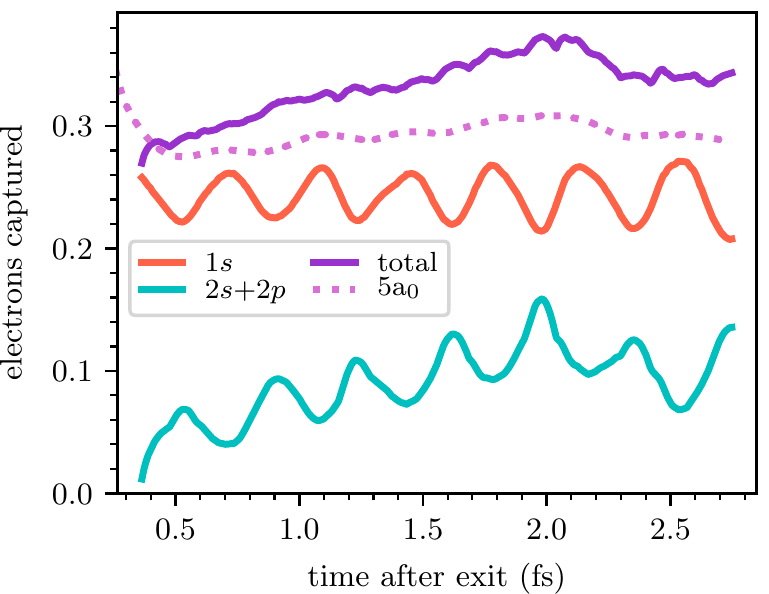}
\caption{
\label{fig:Hocc}(Color online.)
Time-dependent hydrogen orbital occupations after a proton projectile with 1\,at.\,u.\ of velocity traverses a 4-layer aluminum sheet.
The number of electrons within 5\,a$_0$ of the projectile is included for reference.
}
\end{figure}

First, we found that charge captured by the projectile remains nearly constant as a function of aluminum sheet thickness (see Fig.\ \ref{fig:PE_vs_vel_thick}(a)), but decreases with higher proton velocity across the entire velocity range considered (see Fig.\ \ref{fig:PE_vs_vel_thick}(b)).
The majority of the captured electron density occupies the $1s$ orbital, with a smaller portion occupying the $2s$ and $2p$ orbitals.
Figure \ref{fig:PE_vs_vel_thick}(b) also shows that the $2s$ and $2p$ orbital occupation is largely independent of the velocity and the change in total captured charge can be attributed to the 1$s$ shell of the projectile.
Hence, while slower projectiles capture more charge from the target and are close to their 1$s$ ground state, faster projectiles capture less charge but any captured electrons are more likely to occupy an excited state.

The lack of dependence of charge capture on sheet thickness (see Fig.\ \ref{fig:PE_vs_vel_thick}(a)) is surprising given the pre-equilibrium stopping presented in Section \ref{sec:results_energy}.
In particular, this behavior is a departure from experiments employing heavier/higher charge projectiles which observed that charge loss distributions depend on foil thickness\cite{Burgi:1993,Hattass:1999}.
Thus, our results indicate that light ions can experience pre-equilibrium stopping power even after projectile charge equilibration.
This conclusion is further supported by the finding that the projectile's charge almost reaches equilibrium even within the thinnest sheet studied here (see Fig.\ \ref{fig:DDEC}) despite pre-equilibrium stopping power even for the thickest sheets (see inset of Fig.\ \ref{fig:stopping_bulk_SRIM}).
Furthermore, this supports our interpretation that larger pre-equilibrium stopping for light ions is related to the target material, via its surface plasmons, rather than being a property of the projectile charge state.

We also note another interesting surface effect:
The projectile emerges with a higher exit charge state than its effective charge state within the sample.
Our data shows this for a proton with a velocity of 1\,at.\,u., which equilibrates to a charge of about $0.5$ \emph{within} the 8-layer sheet (see Fig.\ \ref{fig:DDEC}), but retains only about $0.3$ electrons when it \emph{exits} with a charge of $0.7$ (see Fig.\ \ref{fig:PE_vs_vel_thick}).
These findings indicate that the exit-side surface strips a portion of electrons that had been captured within the material, an effect not described by existing models of projectile charge equilibration \cite{Brandt:1973,Hattass:1999}.
We find that the number of electrons that are stripped at the surface increases with proton velocity (see Fig.\ \ref{fig:q_stripped}).
These discrepancies between the projectile's charge state within the material and after exiting into the vacuum lead us to advise caution in drawing conclusions about pre-equilibrium behavior from measurements of only the projectile's exit charge state.

Finally, our analysis of the femtosecond dynamics of projectile orbital occupations shows that for \mbox{$v \gtrsim 0.75$\,at.\,u.,} the captured electrons fluctuate between the $1s$ and $2s$/$2p$ states.
We show this explicitly for $v=1.0$\,at.\,u.\ in Fig.\ \ref{fig:Hocc}, and the oscillatory behavior for $v=1.25$\,at.\,u.\ and $1.5$\,at.\,u.\ is similar (see Fig.\ \ref{fig:Hocc_150} of the supplemental material).
The Fourier transform of the time-dependent $1s$ occupation number features a strong peak at a frequency ranging from $\hbar\omega = 10.5 \pm 1.3$\,eV to $11.7\pm 1.2$\,eV, depending on projectile velocity and sheet thickness.
Since these oscillation frequencies lie near the 10.2\,eV energy difference between the $n=1$ and $n=2$ hydrogen orbitals, the fluctuations with these frequencies suggest a superposition of these two states.
We also note that for a projectile velocity of $v=0.75$\,at.\,u., the amplitude of the oscillations in the orbital occupation dynamics becomes very small, making them hard to interpret, and at lower velocities the oscillations disappear (see Fig.\ \ref{fig:Hocc_050} of the supplemental material).
In this low velocity regime, occupation of the $1s$ orbital dominates.
This again indicates that slower projectiles are much closer to their electronic ground state when exiting the aluminum sheet.

\section{\label{sec:conclusion}Conclusions}

Our first-principles simulations of proton-irradiated aluminum sheets revealed detailed information about pre-equilibrium stopping power near a metal's surface.
We found higher average stopping power in the aluminum sheets compared to bulk aluminum and a pronounced stopping power enhancement within the entrance layers.
These deviations from bulk behavior are not adequately explained by pre-equilibrium projectile charge, surface polarization, or quantum confinement; the most viable mechanism for surface stopping enhancement is surface plasmon excitations.

We also presented a novel technique based on analytical hydrogen orbitals to extract from the electron density not only the exit charge state of the projectile, but also the orbital occupations of the captured electrons.
The electrons captured by the proton predominantly occupy the $1s$ orbital, though for higher velocities the projectile emerges in a superposition of $1s$ and $2s$/$2p$ orbitals.
Moreover, the projectile's exit charge state differs from its equilibrium charge within the material, and the number of electrons stripped from the projectile as it emerges from the exit-side surface increases with projectile velocity.

This work provides new details about pre-equilibrium behavior in ion-irradiated surfaces and thin materials, offering unprecedented insight into the few-fs dynamics of electronic stopping power and projectile charge equilibration.
This study also has broad implications for advancing radiation hardness and ion beam techniques for imaging, defect engineering, and nanostructuring, as these applications are chiefly concerned with energy deposition near material surfaces.
The electron density analysis framework developed in this work lays the foundation for further computational research on charge dynamics near ion-irradiated surfaces and 2D materials.

\begin{acknowledgments}
This material is based upon work supported by the National Science Foundation under Grant No.\ OAC-1740219.
Support from the IAEA F11020 CRP "Ion Beam Induced Spatio-temporal Structural Evolution of Materials: Accelerators for a New Technology Era" is gratefully acknowledged.
This research is part of the Blue Waters sustained-petascale computing project, which is supported by the National Science Foundation (awards OCI-0725070 and ACI-1238993) and the state of Illinois.
Blue Waters is a joint effort of the University of Illinois at Urbana-Champaign and its National Center for Supercomputing Applications.
An award of computer time was provided by the Innovative and Novel Computational Impact on Theory and Experiment (INCITE) program.
This research used resources of the Argonne Leadership Computing Facility, which is a DOE Office of Science User Facility supported under Contract DE-AC02-06CH11357.
This work made use of the Illinois Campus Cluster, a computing resource that is operated by the Illinois Campus Cluster Program (ICCP) in conjunction with the National Center for Supercomputing Applications (NCSA) and which is supported by funds from the University of Illinois at Urbana-Champaign.
\end{acknowledgments}

\clearpage
\title{Supplemental Material: Non-adiabatic electron-ion dynamics in proton irradiated aluminum sheets}
\author{Alina Kononov}
\email{kononov2@illinois.edu}
\author{Andr\'{e} Schleife}
\email{schleife@illinois.edu}
\affiliation{University of Illinois at Urbana-Champaign}
\date{\today}

\maketitle

\renewcommand\thefigure{S\arabic{figure}}
\renewcommand{\theHfigure}{S\thefigure}
\setcounter{figure}{0}

\section{Accuracy of time evolution}
\label{sec:integrators}

Significant numerical error may accumulate when integrating the time-dependent Kohn Sham (TDKS) equations for thousands of time steps within the large supercells involved in this study.
To evaluate the accuracy of the time propagation, we measured the error in net charge when evolving a ground state aluminum sheet with no excitation (i.e., without a projectile).
Ideally, such a test simulation should produce a static electron density since the initial wave function is an eigenstate of the time-independent Hamiltonian.

\begin{figure}
\includegraphics{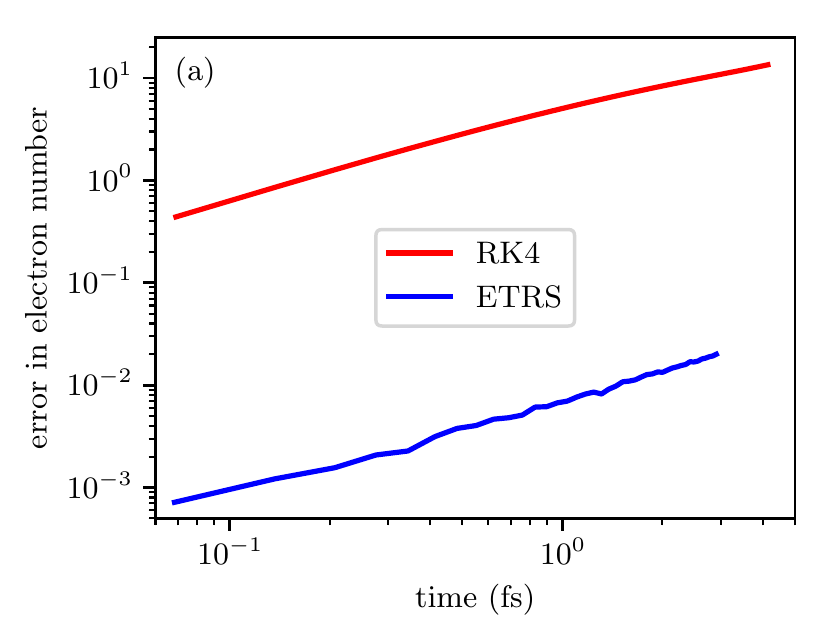}
\includegraphics{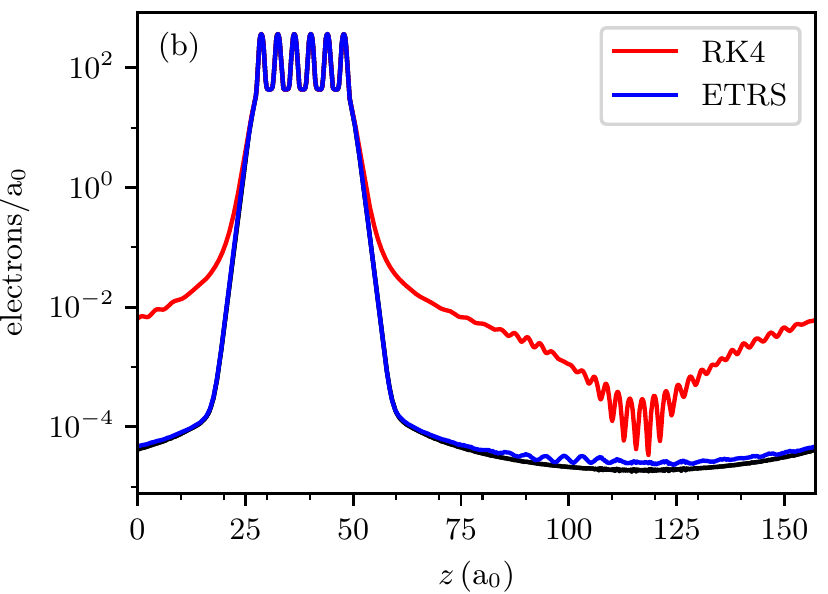}
\caption{(Color online.)
(a) Error in net charge over time when propagating a ground state aluminum sheet (300 atoms) with no excitation using the ETRS and RK4 integrators with time steps of 1.04 and 0.35\,as, respectively.
(b) Electron density profiles from the same test simulations after propagating for 2.9\,fs. The density profile of the initial ground state is shown in black. RK4 and ETRS accumulate about 0.5 and 0.005 fictitious electrons in the vacuum region, respectively.}
\label{fig:integrators}
\end{figure}

We found that the popular fourth-order Runge-Kutta (RK4) propagator \cite{Schleife:2012} accumulates significant numerical errors in net charge (see Fig.\ \ref{fig:integrators}(a)), even when using a small time step of $dt_\mathrm{RK4}=0.35$\,as.
This time step was shown in Ref.\ \onlinecite{Schleife:2015} to be sufficiently accurate in stopping power calculations for light ions traversing bulk aluminum. 
However, accurately simulating 
an ion's interaction with a surface requires much larger supercells containing sufficient vacuum to mitigate finite-size effects. 
Unfortunately, a non-negligible portion of the large error in net charge manifests as a fictitious density in the vacuum region (see Fig.\ \ref{fig:integrators}(b)), which would distort calculations of electron 
dynamics outside the material.

In contrast, we found that the enforced time-reversal symmetry (ETRS) propagator \cite{Draeger:2017} is far more accurate than RK4 without increasing computational cost.
Since each step of the ETRS implementation involves twelve $H\psi$ evaluations compared to RK4's four $H\psi$ evaluations, we retain a similar wall time per simulation time by choosing a time step of $dt_\mathrm{ETRS}=3dt_\mathrm{RK4}=1.04$\,as for ETRS simulations.
Even with this larger time step, we find that ETRS still produces much smaller numerical errors and essentially preserves the ground state density during the test simulations (see Fig.\ \ref{fig:integrators}).
ETRS has also been shown to outperform more advanced Runge-Kutta methods when applied to very similar TDKS problems \cite{Kang:2019}.
Thus, we used the ETRS integrator with a time step of 1.04\,as to produce almost all of the results reported in the main text.
The only exception is the stopping power data point for the 10-layer sheet, which was calculated using RK4 and a time step of 0.35\,as; we find that this choice does not influence stopping power, and we do not analyze the subsequent charge dynamics for this case.
During the simulations for which the electron density is analyzed in the main text, the error in net charge never exceeded 0.04 electrons.

Finally, we confirmed the assumption that atomic motion is negligible over the fs time-scale of the simulations by comparing results from two simulations of a 25\,keV (1\,at.\,u.\ of velocity) H$^+$ projectile impacting a 2-bilayer aluminum sheet: one with fixed ionic velocities and one with true Ehrenfest molecular dynamics.
Allowing the nuclei to respond to Hellmann-Feynman forces changes the
average stopping power by less than 0.02\% and the total charge captured by the projectile by less than 1\%.


\section{\label{sec:cellsize}Convergence of Supercell Dimensions}

\begin{figure}
\centering
\includegraphics{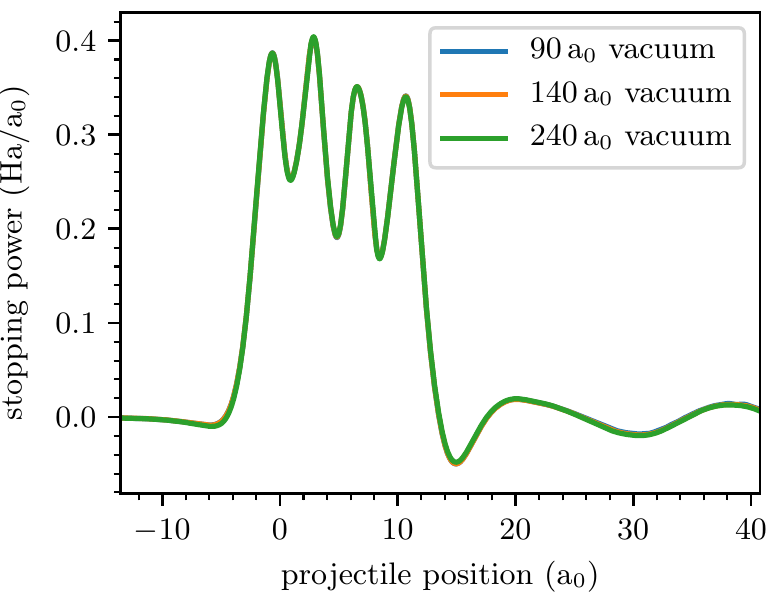}
\caption{(Color online.)
Instantaneous stopping power as a channeling proton with 1\,at.\,u. of velocity impacts a 4-layer aluminum sheet is not sensitive to the vacuum length. The entrance layer of aluminum atoms is located at 0\,a$_0$.}
\label{fig:al_vac_stopping}
\end{figure}

As shown in Fig.\ \ref{fig:al_vac_stopping}, we find that electronic stopping power does not depend strongly on the vacuum length, which we define as the difference between the length of the supercell and the thickness of the material (not accounting for finite surface widths). 
For channeling protons with 1\,at.\,u.\ of velocity, the average stopping power varies by less than 0.2\% within 4-layer aluminum sheets among the different vacuum lengths tested.

\begin{figure}
\centering
\includegraphics{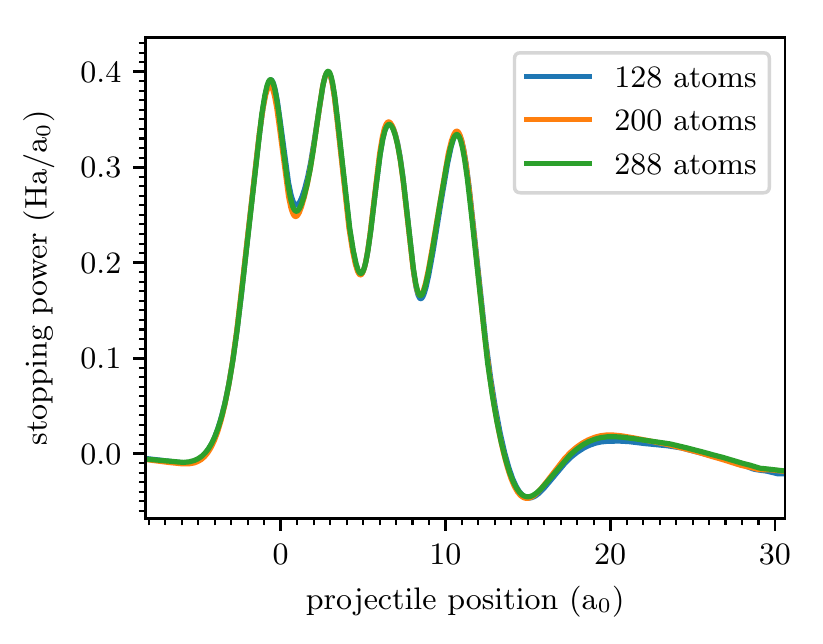}
\caption{(Color online.)
Instantaneous stopping power as a proton with 1\,at.\,u. of velocity traverses a 4-layer aluminum sheet does not change as the lateral dimensions of the sheet are varied.
The entrance layer of aluminum atoms is located at 0\,a$_0$.}
\label{fig:al_sheetsize_stopping}
\end{figure}

To ensure convergence with respect to lateral material dimensions, we simulated protons with 1\,at.\,u.\ of velocity irradiating 4-layer aluminum sheets of area $30.6^2$\,a$_0^2$ (128 atoms), $38.3^2$\,a$_0^2$ (200 atoms), and $45.9^2$\,a$_0^2$ (288 atoms).
Because of the high computational cost of the largest sheet, the vacuum length was reduced to 70\,a$_0$ for these tests.
Also, instead of ETRS with a time step of 0.043\,at.\,u. as decided in Section\ \ref{sec:integrators}, these aluminum calculations used the RK4 time-stepper with a time step of 0.014\,at.\,u.
We do not expect these parameter choices to significantly influence the converged supercell dimensions.
We found very good convergence of the instantaneous stopping power with respect to the area of the material (see Fig.\ \ref{fig:al_sheetsize_stopping}), with average stopping power within the material differing by only 0.5\% among the different size aluminum samples tested.

\section{Charge Capture}
\label{sec:charge_supp}

\begin{figure}
\includegraphics{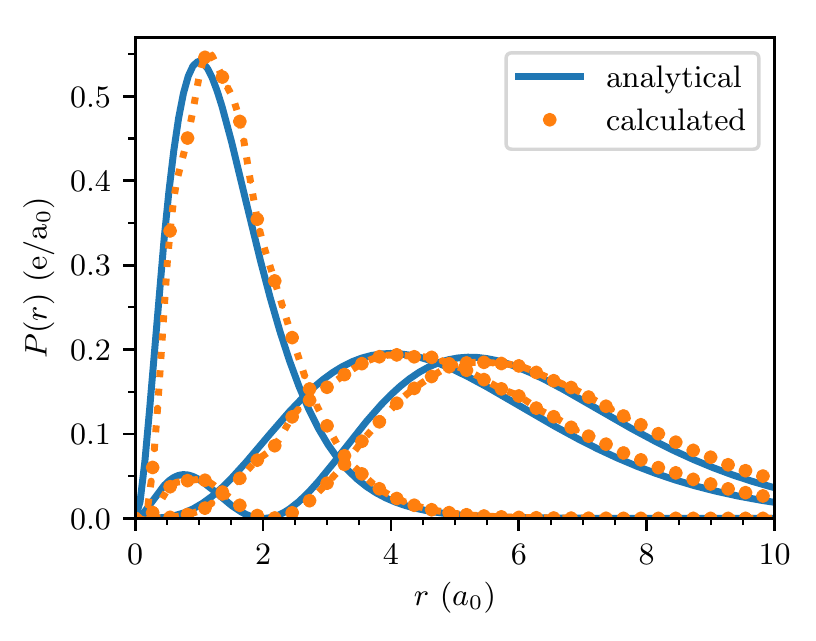}
\caption{(Color online.)
Radial distributions of the analytical hydrogen orbitals and the ground-sate Kohn-Sham orbitals of an isolated H$^+$ ion.
}
\label{fig:H_orbitals}
\end{figure}

First, we validated the applicability of our novel approach for calculating the projectile's exit charge state (see Section \ref{sec:density_analysis}) by examining the hydrogen orbitals calculated by DFT.
We found that the radial distributions of the densities corresponding to the ground-state Kohn-Sham orbitals $\phi_j(\mathbf{r})$ of an isolated H$^+$ ion, i.e.,
$$P_j(r) = \int_0^{2\pi} \int_0^\pi |\phi_j(\mathbf{r})|^2 \; r^2\sin\theta \; d\theta \; d\phi,$$ 
match extremely well with the analytical hydrogen orbitals (see Fig.\ \ref{fig:H_orbitals}).
This suggests that the electrons captured by the proton projectile in our TDDFT simulations can be adequately described with analytical hydrogen orbitals.
Also, for heavier ions where analytical orbitals are not available,  our novel approach can still be applied using calculated radial distributions of the isolated projectile.

\begin{figure}
\includegraphics{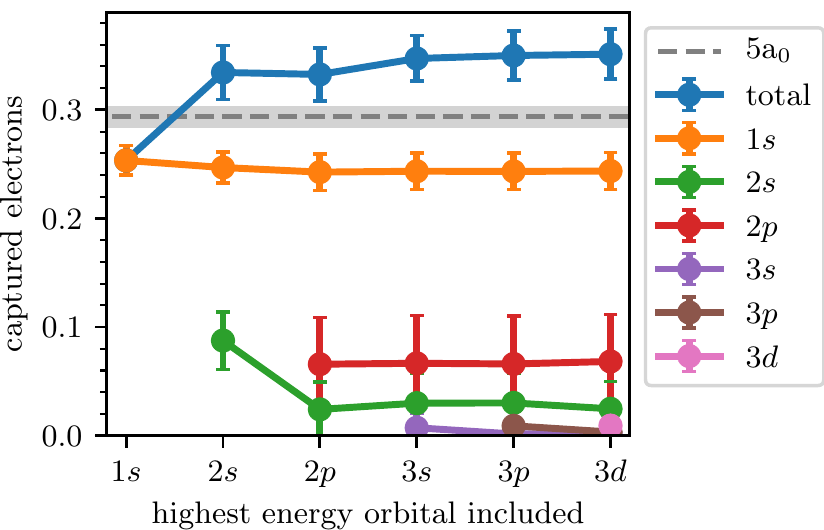}
\caption{(Color online.)
Projectile orbital occupations and total captured electrons obtained from the novel orbital fitting technique (see Section \ref{sec:density_analysis}) as the number of included orbitals varies. The time-dependent occupations are averaged over time after a proton with velocity 1\,at.\,u\ traverses a 4-layer aluminum sheet, and their standard deviations are illustrated by error bars. The average number of electrons within 5\,a$_0$ of the projectile, shaded to indicate standard deviation, is included for reference.}
\label{fig:moreorbitals}
\end{figure}

We also found that including higher energy orbitals (above $2p$) in the orbital fitting does not significantly alter the orbital occupations or the total charge capture (see Fig.\ \ref{fig:moreorbitals}).
Therefore, we concluded that contributions from orbitals above $n=2$ are negligible and we considered only the $1s$, $2s$, and $2p$ orbitals when computing the charge capture results presented in the main text.

\begin{figure}
\includegraphics{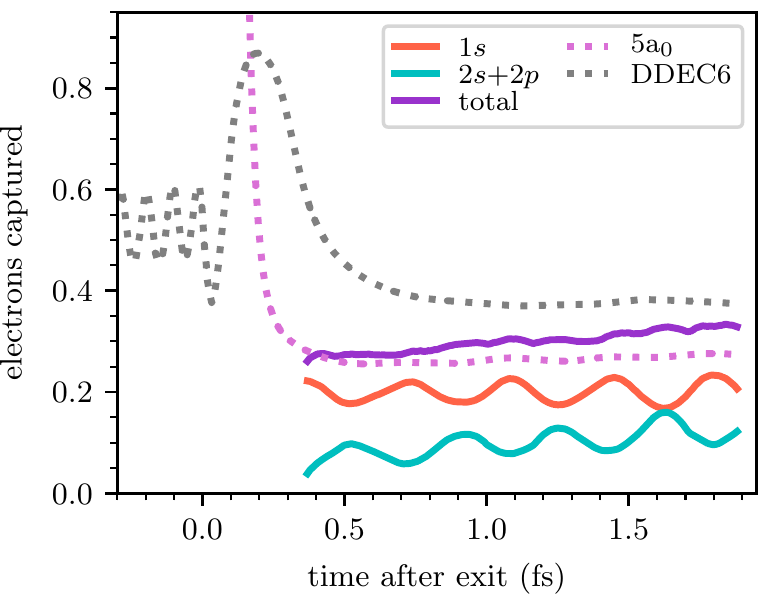}
\caption{
\label{fig:DDEC_vs_fit}(Color online.)
Charge captured by a H$^+$ projectile with 1.0\,at.\,u.\ of velocity after traversing an 8-layer aluminum sheet.
Results computed by DDEC6 analysis \cite{Manz:2016,Gabaldon_limas:2016} are compared to the novel fitting technique (see Section \ref{sec:density_analysis}) and the number of electrons within 5\,a$_0$ of the projectile.
}
\end{figure}

While the simple volume partitioning method of integrating the electron density within a 5\,a$_0$ radius of the projectile produces qualitatively similar results for the total charge captured as our novel approach, the DDEC6 charge partitioning method overestimates the total charge captured compared to the other two techniques (see Fig.\ \ref{fig:DDEC_vs_fit}).
We attribute this overestimation to this method's inapplicability in the presence of a free electron distribution, which causes erroneous assignment of nearby emitted electrons to the projectile.
We found that the DDEC6 results are particularly inaccurate when the projectile is still near the exit-side surface, but become qualitatively similar to the other methods over time.

\begin{figure}
\includegraphics{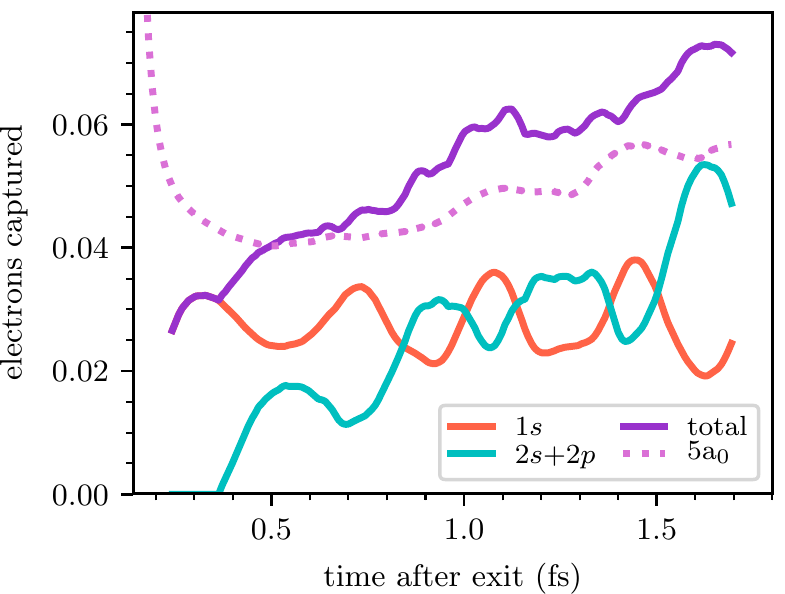}
\caption{
\label{fig:Hocc_150}(Color online.)
Time-dependent hydrogen orbital occupations after a H$^+$ projectile with a velocity of 1.5\,at.\,u.\ traverses a 4-layer aluminum sheet.
The number of electrons within 5\,a$_0$ of the projectile is included for reference.
}
\end{figure}

\begin{figure}
\includegraphics{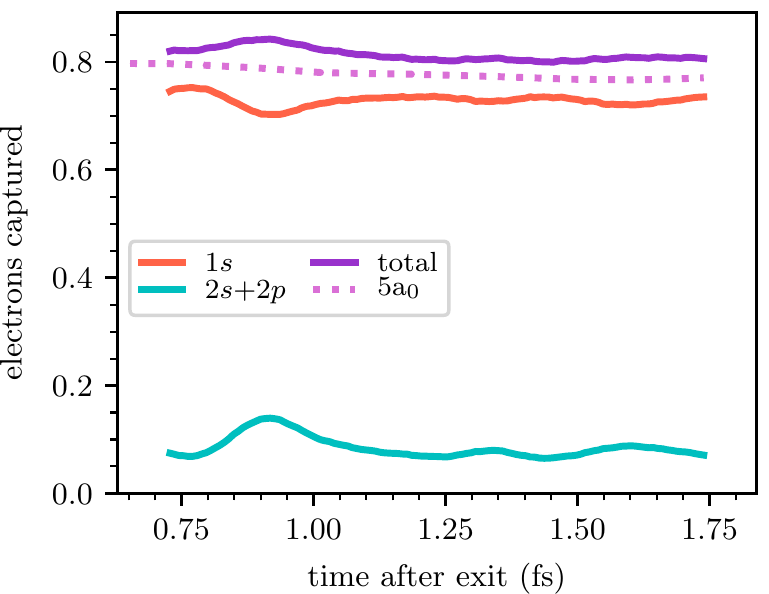}
\caption{
\label{fig:Hocc_050}(Color online.)
Time-dependent hydrogen orbital occupations after a H$^+$ projectile with a velocity of 0.5\,at.\,u.\ traverses a 4-layer aluminum sheet.
The number of electrons within 5\,a$_0$ of the projectile is included for reference.
}
\end{figure}

Finally, we present additional results for projectile orbital occupation dynamics.
Figure \ref{fig:Hocc_150} illustrates similar oscillatory behavior at $v=1.5$\,at.\,u.\ as was shown for $v=1.0$\,at.\,u.\ in Fig.\ \ref{fig:Hocc}.
However, for proton velocities below $\sim 0.75$\,at.\,u., these oscillations disappear and occupation of the $1s$ orbital dominates (see Fig.\ \ref{fig:Hocc_050}).

\bibliography{main.bib}

\begin{thebibliography}{89}%
\makeatletter
\providecommand \@ifxundefined [1]{%
 \@ifx{#1\undefined}
}%
\providecommand \@ifnum [1]{%
 \ifnum #1\expandafter \@firstoftwo
 \else \expandafter \@secondoftwo
 \fi
}%
\providecommand \@ifx [1]{%
 \ifx #1\expandafter \@firstoftwo
 \else \expandafter \@secondoftwo
 \fi
}%
\providecommand \natexlab [1]{#1}%
\providecommand \enquote  [1]{``#1''}%
\providecommand \bibnamefont  [1]{#1}%
\providecommand \bibfnamefont [1]{#1}%
\providecommand \citenamefont [1]{#1}%
\providecommand \href@noop [0]{\@secondoftwo}%
\providecommand \href [0]{\begingroup \@sanitize@url \@href}%
\providecommand \@href[1]{\@@startlink{#1}\@@href}%
\providecommand \@@href[1]{\endgroup#1\@@endlink}%
\providecommand \@sanitize@url [0]{\catcode `\\12\catcode `\$12\catcode
  `\&12\catcode `\#12\catcode `\^12\catcode `\_12\catcode `\%12\relax}%
\providecommand \@@startlink[1]{}%
\providecommand \@@endlink[0]{}%
\providecommand \url  [0]{\begingroup\@sanitize@url \@url }%
\providecommand \@url [1]{\endgroup\@href {#1}{\urlprefix }}%
\providecommand \urlprefix  [0]{URL }%
\providecommand \Eprint [0]{\href }%
\providecommand \doibase [0]{http://dx.doi.org/}%
\providecommand \selectlanguage [0]{\@gobble}%
\providecommand \bibinfo  [0]{\@secondoftwo}%
\providecommand \bibfield  [0]{\@secondoftwo}%
\providecommand \translation [1]{[#1]}%
\providecommand \BibitemOpen [0]{}%
\providecommand \bibitemStop [0]{}%
\providecommand \bibitemNoStop [0]{.\EOS\space}%
\providecommand \EOS [0]{\spacefactor3000\relax}%
\providecommand \BibitemShut  [1]{\csname bibitem#1\endcsname}%
\let\auto@bib@innerbib\@empty
\bibitem [{\citenamefont {Notte}\ \emph {et~al.}(2007)\citenamefont {Notte},
  \citenamefont {Ward}, \citenamefont {Economou}, \citenamefont {Hill},
  \citenamefont {Percival}, \citenamefont {Farkas},\ and\ \citenamefont
  {McVey}}]{Notte:2007}%
  \BibitemOpen
  \bibfield  {author} {\bibinfo {author} {\bibfnamefont {J.}~\bibnamefont
  {Notte}}, \bibinfo {author} {\bibfnamefont {B.}~\bibnamefont {Ward}},
  \bibinfo {author} {\bibfnamefont {N.}~\bibnamefont {Economou}}, \bibinfo
  {author} {\bibfnamefont {R.}~\bibnamefont {Hill}}, \bibinfo {author}
  {\bibfnamefont {R.}~\bibnamefont {Percival}}, \bibinfo {author}
  {\bibfnamefont {L.}~\bibnamefont {Farkas}}, \ and\ \bibinfo {author}
  {\bibfnamefont {S.}~\bibnamefont {McVey}},\ }\href {\doibase
  10.1063/1.2799423} {\bibfield  {journal} {\bibinfo  {journal} {AIP Conference
  Proceedings}\ }\textbf {\bibinfo {volume} {931}},\ \bibinfo {pages} {489}
  (\bibinfo {year} {2007})}\BibitemShut {NoStop}%
\bibitem [{\citenamefont {Utke}\ \emph {et~al.}(2008)\citenamefont {Utke},
  \citenamefont {Hoffmann},\ and\ \citenamefont {Melngailis}}]{Utke:2008}%
  \BibitemOpen
  \bibfield  {author} {\bibinfo {author} {\bibfnamefont {I.}~\bibnamefont
  {Utke}}, \bibinfo {author} {\bibfnamefont {P.}~\bibnamefont {Hoffmann}}, \
  and\ \bibinfo {author} {\bibfnamefont {J.}~\bibnamefont {Melngailis}},\
  }\href {\doibase 10.1116/1.2955728} {\bibfield  {journal} {\bibinfo
  {journal} {Journal of Vacuum Science \& Technology B: Microelectronics and
  Nanometer Structures}\ }\textbf {\bibinfo {volume} {26}},\ \bibinfo {pages}
  {1197} (\bibinfo {year} {2008})}\BibitemShut {NoStop}%
\bibitem [{\citenamefont {Ritter}\ \emph {et~al.}(2013)\citenamefont {Ritter},
  \citenamefont {Wilhelm}, \citenamefont {Stöger-Pollach}, \citenamefont
  {Heller}, \citenamefont {Mücklich}, \citenamefont {Werner}, \citenamefont
  {Vieker}, \citenamefont {Beyer}, \citenamefont {Facsko}, \citenamefont
  {Gölzhäuser},\ and\ \citenamefont {Aumayr}}]{Ritter:2013}%
  \BibitemOpen
  \bibfield  {author} {\bibinfo {author} {\bibfnamefont {R.}~\bibnamefont
  {Ritter}}, \bibinfo {author} {\bibfnamefont {R.~A.}\ \bibnamefont {Wilhelm}},
  \bibinfo {author} {\bibfnamefont {M.}~\bibnamefont {Stöger-Pollach}},
  \bibinfo {author} {\bibfnamefont {R.}~\bibnamefont {Heller}}, \bibinfo
  {author} {\bibfnamefont {A.}~\bibnamefont {Mücklich}}, \bibinfo {author}
  {\bibfnamefont {U.}~\bibnamefont {Werner}}, \bibinfo {author} {\bibfnamefont
  {H.}~\bibnamefont {Vieker}}, \bibinfo {author} {\bibfnamefont
  {A.}~\bibnamefont {Beyer}}, \bibinfo {author} {\bibfnamefont
  {S.}~\bibnamefont {Facsko}}, \bibinfo {author} {\bibfnamefont
  {A.}~\bibnamefont {Gölzhäuser}}, \ and\ \bibinfo {author} {\bibfnamefont
  {F.}~\bibnamefont {Aumayr}},\ }\href {\doibase 10.1063/1.4792511} {\bibfield
  {journal} {\bibinfo  {journal} {Applied Physics Letters}\ }\textbf {\bibinfo
  {volume} {102}},\ \bibinfo {pages} {063112} (\bibinfo {year}
  {2013})}\BibitemShut {NoStop}%
\bibitem [{\citenamefont {Hlawacek}\ \emph {et~al.}(2014)\citenamefont
  {Hlawacek}, \citenamefont {Veligura}, \citenamefont {van Gastel},\ and\
  \citenamefont {Poelsema}}]{Hlawacek:2014}%
  \BibitemOpen
  \bibfield  {author} {\bibinfo {author} {\bibfnamefont {G.}~\bibnamefont
  {Hlawacek}}, \bibinfo {author} {\bibfnamefont {V.}~\bibnamefont {Veligura}},
  \bibinfo {author} {\bibfnamefont {R.}~\bibnamefont {van Gastel}}, \ and\
  \bibinfo {author} {\bibfnamefont {B.}~\bibnamefont {Poelsema}},\ }\href
  {\doibase 10.1116/1.4863676} {\bibfield  {journal} {\bibinfo  {journal} {J.
  Vac. Sci. Technol. B}\ }\textbf {\bibinfo {volume} {32}},\ \bibinfo {eid}
  {020801} (\bibinfo {year} {2014}),\ 10.1116/1.4863676}\BibitemShut {NoStop}%
\bibitem [{\citenamefont {Fox}\ \emph {et~al.}(2015)\citenamefont {Fox},
  \citenamefont {Zhou}, \citenamefont {Maguire}, \citenamefont {O’Neill},
  \citenamefont {Ó’Coileáin}, \citenamefont {Gatensby}, \citenamefont
  {Glushenkov}, \citenamefont {Tao}, \citenamefont {Duesberg}, \citenamefont
  {Shvets}, \citenamefont {Abid}, \citenamefont {Abid}, \citenamefont {Wu},
  \citenamefont {Chen}, \citenamefont {Coleman}, \citenamefont {Donegan},\ and\
  \citenamefont {Zhang}}]{Fox:2015}%
  \BibitemOpen
  \bibfield  {author} {\bibinfo {author} {\bibfnamefont {D.~S.}\ \bibnamefont
  {Fox}}, \bibinfo {author} {\bibfnamefont {Y.}~\bibnamefont {Zhou}}, \bibinfo
  {author} {\bibfnamefont {P.}~\bibnamefont {Maguire}}, \bibinfo {author}
  {\bibfnamefont {A.}~\bibnamefont {O’Neill}}, \bibinfo {author}
  {\bibfnamefont {C.}~\bibnamefont {Ó’Coileáin}}, \bibinfo {author}
  {\bibfnamefont {R.}~\bibnamefont {Gatensby}}, \bibinfo {author}
  {\bibfnamefont {A.~M.}\ \bibnamefont {Glushenkov}}, \bibinfo {author}
  {\bibfnamefont {T.}~\bibnamefont {Tao}}, \bibinfo {author} {\bibfnamefont
  {G.~S.}\ \bibnamefont {Duesberg}}, \bibinfo {author} {\bibfnamefont {I.~V.}\
  \bibnamefont {Shvets}}, \bibinfo {author} {\bibfnamefont {M.}~\bibnamefont
  {Abid}}, \bibinfo {author} {\bibfnamefont {M.}~\bibnamefont {Abid}}, \bibinfo
  {author} {\bibfnamefont {H.-C.}\ \bibnamefont {Wu}}, \bibinfo {author}
  {\bibfnamefont {Y.}~\bibnamefont {Chen}}, \bibinfo {author} {\bibfnamefont
  {J.~N.}\ \bibnamefont {Coleman}}, \bibinfo {author} {\bibfnamefont {J.~F.}\
  \bibnamefont {Donegan}}, \ and\ \bibinfo {author} {\bibfnamefont
  {H.}~\bibnamefont {Zhang}},\ }\href {\doibase 10.1021/acs.nanolett.5b01673}
  {\bibfield  {journal} {\bibinfo  {journal} {Nano Letters}\ }\textbf {\bibinfo
  {volume} {15}},\ \bibinfo {pages} {5307} (\bibinfo {year}
  {2015})}\BibitemShut {NoStop}%
\bibitem [{\citenamefont {Li}\ and\ \citenamefont {Chen}(2017)}]{Li:2017:2D}%
  \BibitemOpen
  \bibfield  {author} {\bibinfo {author} {\bibfnamefont {Z.}~\bibnamefont
  {Li}}\ and\ \bibinfo {author} {\bibfnamefont {F.}~\bibnamefont {Chen}},\
  }\href {\doibase 10.1063/1.4977087} {\bibfield  {journal} {\bibinfo
  {journal} {Applied Physics Reviews}\ }\textbf {\bibinfo {volume} {4}},\
  \bibinfo {pages} {011103} (\bibinfo {year} {2017})}\BibitemShut {NoStop}%
\bibitem [{\citenamefont {Bethe}(1930)}]{Bethe:1930}%
  \BibitemOpen
  \bibfield  {author} {\bibinfo {author} {\bibfnamefont {H.}~\bibnamefont
  {Bethe}},\ }\href {\doibase 10.1002/andp.19303970303} {\bibfield  {journal}
  {\bibinfo  {journal} {Annalen der Physik}\ }\textbf {\bibinfo {volume}
  {397}},\ \bibinfo {pages} {325} (\bibinfo {year} {1930})}\BibitemShut
  {NoStop}%
\bibitem [{\citenamefont {Sigmund}(2008)}]{Sigmund:2008}%
  \BibitemOpen
  \bibfield  {author} {\bibinfo {author} {\bibfnamefont {P.}~\bibnamefont
  {Sigmund}},\ }\href@noop {} {\emph {\bibinfo {title} {Particle penetration
  and radiation effects: general aspects and stopping of swift point
  charges}}},\ \bibinfo {series} {Springer series in solid-state science}\ No.\
  \bibinfo {number} {151}\ (\bibinfo  {publisher} {Springer},\ \bibinfo {year}
  {2008})\BibitemShut {NoStop}%
\bibitem [{\citenamefont {Fermi}\ and\ \citenamefont
  {Teller}(1947)}]{Fermi:1947}%
  \BibitemOpen
  \bibfield  {author} {\bibinfo {author} {\bibfnamefont {E.}~\bibnamefont
  {Fermi}}\ and\ \bibinfo {author} {\bibfnamefont {E.}~\bibnamefont {Teller}},\
  }\href {\doibase 10.1103/PhysRev.72.399} {\bibfield  {journal} {\bibinfo
  {journal} {Physical Review}\ }\textbf {\bibinfo {volume} {72}},\ \bibinfo
  {pages} {399} (\bibinfo {year} {1947})}\BibitemShut {NoStop}%
\bibitem [{\citenamefont {Lindhard}\ and\ \citenamefont
  {Winther}(1964)}]{Lindhard:1964}%
  \BibitemOpen
  \bibfield  {author} {\bibinfo {author} {\bibfnamefont {J.}~\bibnamefont
  {Lindhard}}\ and\ \bibinfo {author} {\bibfnamefont {A.}~\bibnamefont
  {Winther}},\ }\href@noop {} {\bibfield  {journal} {\bibinfo  {journal}
  {Matematisk-fysiske Meddelelser}\ }\textbf {\bibinfo {volume} {34}},\
  \bibinfo {pages} {1} (\bibinfo {year} {1964})}\BibitemShut {NoStop}%
\bibitem [{\citenamefont {Wang}\ \emph {et~al.}(1998)\citenamefont {Wang},
  \citenamefont {Mehlhorn},\ and\ \citenamefont {{MacFarlane}}}]{Wang:1998}%
  \BibitemOpen
  \bibfield  {author} {\bibinfo {author} {\bibfnamefont {P.}~\bibnamefont
  {Wang}}, \bibinfo {author} {\bibfnamefont {T.~M.}\ \bibnamefont {Mehlhorn}},
  \ and\ \bibinfo {author} {\bibfnamefont {J.~J.}\ \bibnamefont
  {{MacFarlane}}},\ }\href {\doibase 10.1063/1.873022} {\bibfield  {journal}
  {\bibinfo  {journal} {Physics of Plasmas}\ }\textbf {\bibinfo {volume} {5}},\
  \bibinfo {pages} {2977} (\bibinfo {year} {1998})}\BibitemShut {NoStop}%
\bibitem [{\citenamefont {Correa}(2018)}]{Correa:2018}%
  \BibitemOpen
  \bibfield  {author} {\bibinfo {author} {\bibfnamefont {A.~A.}\ \bibnamefont
  {Correa}},\ }\href {\doibase 10.1016/j.commatsci.2018.03.064} {\bibfield
  {journal} {\bibinfo  {journal} {Computational Materials Science}\ }\textbf
  {\bibinfo {volume} {150}},\ \bibinfo {pages} {291} (\bibinfo {year}
  {2018})}\BibitemShut {NoStop}%
\bibitem [{\citenamefont {Lee}\ \emph {et~al.}()\citenamefont {Lee},
  \citenamefont {Stewart}, \citenamefont {Foiles}, \citenamefont
  {Dingreville},\ and\ \citenamefont {Schleife}}]{Lee:2020}%
  \BibitemOpen
  \bibfield  {author} {\bibinfo {author} {\bibfnamefont {C.-W.}\ \bibnamefont
  {Lee}}, \bibinfo {author} {\bibfnamefont {J.~A.}\ \bibnamefont {Stewart}},
  \bibinfo {author} {\bibfnamefont {S.~M.}\ \bibnamefont {Foiles}}, \bibinfo
  {author} {\bibfnamefont {R.}~\bibnamefont {Dingreville}}, \ and\ \bibinfo
  {author} {\bibfnamefont {A.}~\bibnamefont {Schleife}},\ }\href
  {http://arxiv.org/abs/2001.10162} {\bibfield  {journal} {\bibinfo  {journal}
  {{arXiv}:2001.10162 [cond-mat]}\ }}\Eprint {http://arxiv.org/abs/2001.10162}
  {2001.10162} \BibitemShut {NoStop}%
\bibitem [{\citenamefont {Hattass}\ \emph {et~al.}(1999)\citenamefont
  {Hattass}, \citenamefont {Schenkel}, \citenamefont {Hamza}, \citenamefont
  {Barnes}, \citenamefont {Newman}, \citenamefont {{McDonald}}, \citenamefont
  {Niedermayr}, \citenamefont {Machicoane},\ and\ \citenamefont
  {Schneider}}]{Hattass:1999}%
  \BibitemOpen
  \bibfield  {author} {\bibinfo {author} {\bibfnamefont {M.}~\bibnamefont
  {Hattass}}, \bibinfo {author} {\bibfnamefont {T.}~\bibnamefont {Schenkel}},
  \bibinfo {author} {\bibfnamefont {A.~V.}\ \bibnamefont {Hamza}}, \bibinfo
  {author} {\bibfnamefont {A.~V.}\ \bibnamefont {Barnes}}, \bibinfo {author}
  {\bibfnamefont {M.~W.}\ \bibnamefont {Newman}}, \bibinfo {author}
  {\bibfnamefont {J.~W.}\ \bibnamefont {{McDonald}}}, \bibinfo {author}
  {\bibfnamefont {T.~R.}\ \bibnamefont {Niedermayr}}, \bibinfo {author}
  {\bibfnamefont {G.~A.}\ \bibnamefont {Machicoane}}, \ and\ \bibinfo {author}
  {\bibfnamefont {D.~H.}\ \bibnamefont {Schneider}},\ }\href {\doibase
  10.1103/PhysRevLett.82.4795} {\bibfield  {journal} {\bibinfo  {journal}
  {Physical Review Letters}\ }\textbf {\bibinfo {volume} {82}},\ \bibinfo
  {pages} {4795} (\bibinfo {year} {1999})}\BibitemShut {NoStop}%
\bibitem [{\citenamefont {Wilhelm}\ \emph {et~al.}(2014)\citenamefont
  {Wilhelm}, \citenamefont {Gruber}, \citenamefont {Ritter}, \citenamefont
  {Heller}, \citenamefont {Facsko},\ and\ \citenamefont
  {Aumayr}}]{Wilhelm:2014}%
  \BibitemOpen
  \bibfield  {author} {\bibinfo {author} {\bibfnamefont {R.~A.}\ \bibnamefont
  {Wilhelm}}, \bibinfo {author} {\bibfnamefont {E.}~\bibnamefont {Gruber}},
  \bibinfo {author} {\bibfnamefont {R.}~\bibnamefont {Ritter}}, \bibinfo
  {author} {\bibfnamefont {R.}~\bibnamefont {Heller}}, \bibinfo {author}
  {\bibfnamefont {S.}~\bibnamefont {Facsko}}, \ and\ \bibinfo {author}
  {\bibfnamefont {F.}~\bibnamefont {Aumayr}},\ }\href {\doibase
  10.1103/PhysRevLett.112.153201} {\bibfield  {journal} {\bibinfo  {journal}
  {Physical Review Letters}\ }\textbf {\bibinfo {volume} {112}},\ \bibinfo
  {pages} {153201} (\bibinfo {year} {2014})}\BibitemShut {NoStop}%
\bibitem [{\citenamefont {Schenkel}\ \emph {et~al.}(1997)\citenamefont
  {Schenkel}, \citenamefont {Briere}, \citenamefont {Schmidt-Böcking},
  \citenamefont {Bethge},\ and\ \citenamefont {Schneider}}]{Schenkel:1997}%
  \BibitemOpen
  \bibfield  {author} {\bibinfo {author} {\bibfnamefont {T.}~\bibnamefont
  {Schenkel}}, \bibinfo {author} {\bibfnamefont {M.~A.}\ \bibnamefont
  {Briere}}, \bibinfo {author} {\bibfnamefont {H.}~\bibnamefont
  {Schmidt-Böcking}}, \bibinfo {author} {\bibfnamefont {K.}~\bibnamefont
  {Bethge}}, \ and\ \bibinfo {author} {\bibfnamefont {D.~H.}\ \bibnamefont
  {Schneider}},\ }\href {\doibase 10.1103/PhysRevLett.78.2481} {\bibfield
  {journal} {\bibinfo  {journal} {Physical Review Letters}\ }\textbf {\bibinfo
  {volume} {78}},\ \bibinfo {pages} {2481} (\bibinfo {year}
  {1997})}\BibitemShut {NoStop}%
\bibitem [{\citenamefont {Sun}\ \emph {et~al.}(1996)\citenamefont {Sun},
  \citenamefont {Yu}, \citenamefont {Lin}, \citenamefont {Wang}, \citenamefont
  {Duggan}, \citenamefont {Azordegan}, \citenamefont {{McDaniel}},\ and\
  \citenamefont {Lapicki}}]{Sun:1996}%
  \BibitemOpen
  \bibfield  {author} {\bibinfo {author} {\bibfnamefont {H.~L.}\ \bibnamefont
  {Sun}}, \bibinfo {author} {\bibfnamefont {Y.~C.}\ \bibnamefont {Yu}},
  \bibinfo {author} {\bibfnamefont {E.~K.}\ \bibnamefont {Lin}}, \bibinfo
  {author} {\bibfnamefont {C.~W.}\ \bibnamefont {Wang}}, \bibinfo {author}
  {\bibfnamefont {J.~L.}\ \bibnamefont {Duggan}}, \bibinfo {author}
  {\bibfnamefont {A.~R.}\ \bibnamefont {Azordegan}}, \bibinfo {author}
  {\bibfnamefont {F.~D.}\ \bibnamefont {{McDaniel}}}, \ and\ \bibinfo {author}
  {\bibfnamefont {G.}~\bibnamefont {Lapicki}},\ }\href {\doibase
  10.1103/PhysRevA.53.4190} {\bibfield  {journal} {\bibinfo  {journal}
  {Physical Review A}\ }\textbf {\bibinfo {volume} {53}},\ \bibinfo {pages}
  {4190} (\bibinfo {year} {1996})}\BibitemShut {NoStop}%
\bibitem [{\citenamefont {Brandt}\ \emph {et~al.}(1973)\citenamefont {Brandt},
  \citenamefont {Laubert}, \citenamefont {Mourino},\ and\ \citenamefont
  {Schwarzschild}}]{Brandt:1973}%
  \BibitemOpen
  \bibfield  {author} {\bibinfo {author} {\bibfnamefont {W.}~\bibnamefont
  {Brandt}}, \bibinfo {author} {\bibfnamefont {R.}~\bibnamefont {Laubert}},
  \bibinfo {author} {\bibfnamefont {M.}~\bibnamefont {Mourino}}, \ and\
  \bibinfo {author} {\bibfnamefont {A.}~\bibnamefont {Schwarzschild}},\ }\href
  {\doibase 10.1103/PhysRevLett.30.358} {\bibfield  {journal} {\bibinfo
  {journal} {Physical Review Letters}\ }\textbf {\bibinfo {volume} {30}},\
  \bibinfo {pages} {358} (\bibinfo {year} {1973})}\BibitemShut {NoStop}%
\bibitem [{\citenamefont {Wilhelm}\ \emph {et~al.}(2017)\citenamefont
  {Wilhelm}, \citenamefont {Gruber}, \citenamefont {Schwestka}, \citenamefont
  {Kozubek}, \citenamefont {Madeira}, \citenamefont {Marques}, \citenamefont
  {Kobus}, \citenamefont {Krasheninnikov}, \citenamefont {Schleberger},\ and\
  \citenamefont {Aumayr}}]{Wilhelm:2017}%
  \BibitemOpen
  \bibfield  {author} {\bibinfo {author} {\bibfnamefont {R.~A.}\ \bibnamefont
  {Wilhelm}}, \bibinfo {author} {\bibfnamefont {E.}~\bibnamefont {Gruber}},
  \bibinfo {author} {\bibfnamefont {J.}~\bibnamefont {Schwestka}}, \bibinfo
  {author} {\bibfnamefont {R.}~\bibnamefont {Kozubek}}, \bibinfo {author}
  {\bibfnamefont {T.~I.}\ \bibnamefont {Madeira}}, \bibinfo {author}
  {\bibfnamefont {J.~P.}\ \bibnamefont {Marques}}, \bibinfo {author}
  {\bibfnamefont {J.}~\bibnamefont {Kobus}}, \bibinfo {author} {\bibfnamefont
  {A.~V.}\ \bibnamefont {Krasheninnikov}}, \bibinfo {author} {\bibfnamefont
  {M.}~\bibnamefont {Schleberger}}, \ and\ \bibinfo {author} {\bibfnamefont
  {F.}~\bibnamefont {Aumayr}},\ }\href {\doibase
  10.1103/PhysRevLett.119.103401} {\bibfield  {journal} {\bibinfo  {journal}
  {Physical Review Letters}\ }\textbf {\bibinfo {volume} {119}},\ \bibinfo
  {pages} {103401} (\bibinfo {year} {2017})}\BibitemShut {NoStop}%
\bibitem [{\citenamefont {Wilhelm}\ \emph {et~al.}(2018)\citenamefont
  {Wilhelm}, \citenamefont {Gruber}, \citenamefont {Schwestka}, \citenamefont
  {Heller}, \citenamefont {Fascko},\ and\ \citenamefont
  {Aumayr}}]{Wilhelm:2018}%
  \BibitemOpen
  \bibfield  {author} {\bibinfo {author} {\bibfnamefont {R.~A.}\ \bibnamefont
  {Wilhelm}}, \bibinfo {author} {\bibfnamefont {E.}~\bibnamefont {Gruber}},
  \bibinfo {author} {\bibfnamefont {J.}~\bibnamefont {Schwestka}}, \bibinfo
  {author} {\bibfnamefont {R.}~\bibnamefont {Heller}}, \bibinfo {author}
  {\bibfnamefont {S.}~\bibnamefont {Fascko}}, \ and\ \bibinfo {author}
  {\bibfnamefont {F.}~\bibnamefont {Aumayr}},\ }\href {\doibase
  10.3390/app8071050} {\bibfield  {journal} {\bibinfo  {journal} {Applied
  Sciences}\ }\textbf {\bibinfo {volume} {8}},\ \bibinfo {pages} {1050}
  (\bibinfo {year} {2018})}\BibitemShut {NoStop}%
\bibitem [{\citenamefont {Koschar}\ \emph {et~al.}(1989)\citenamefont
  {Koschar}, \citenamefont {Kroneberger}, \citenamefont {Clouvas},
  \citenamefont {Burkhard}, \citenamefont {Meckbach}, \citenamefont {Heil},
  \citenamefont {Kemmler}, \citenamefont {Rothard}, \citenamefont {Groeneveld},
  \citenamefont {Schramm},\ and\ \citenamefont {Betz}}]{Koschar:1989}%
  \BibitemOpen
  \bibfield  {author} {\bibinfo {author} {\bibfnamefont {P.}~\bibnamefont
  {Koschar}}, \bibinfo {author} {\bibfnamefont {K.}~\bibnamefont
  {Kroneberger}}, \bibinfo {author} {\bibfnamefont {A.}~\bibnamefont
  {Clouvas}}, \bibinfo {author} {\bibfnamefont {M.}~\bibnamefont {Burkhard}},
  \bibinfo {author} {\bibfnamefont {W.}~\bibnamefont {Meckbach}}, \bibinfo
  {author} {\bibfnamefont {O.}~\bibnamefont {Heil}}, \bibinfo {author}
  {\bibfnamefont {J.}~\bibnamefont {Kemmler}}, \bibinfo {author} {\bibfnamefont
  {H.}~\bibnamefont {Rothard}}, \bibinfo {author} {\bibfnamefont {K.~O.}\
  \bibnamefont {Groeneveld}}, \bibinfo {author} {\bibfnamefont
  {R.}~\bibnamefont {Schramm}}, \ and\ \bibinfo {author} {\bibfnamefont
  {H.-D.}\ \bibnamefont {Betz}},\ }\href {\doibase 10.1103/PhysRevA.40.3632}
  {\bibfield  {journal} {\bibinfo  {journal} {Physical Review A}\ }\textbf
  {\bibinfo {volume} {40}},\ \bibinfo {pages} {3632} (\bibinfo {year}
  {1989})}\BibitemShut {NoStop}%
\bibitem [{\citenamefont {Mertens}\ and\ \citenamefont
  {Krist}(1982)}]{Mertens:1982}%
  \BibitemOpen
  \bibfield  {author} {\bibinfo {author} {\bibfnamefont {P.}~\bibnamefont
  {Mertens}}\ and\ \bibinfo {author} {\bibfnamefont {T.}~\bibnamefont
  {Krist}},\ }\href {\doibase 10.1016/0029-554X(82)90489-X} {\bibfield
  {journal} {\bibinfo  {journal} {Nuclear Instruments and Methods in Physics
  Research}\ }\textbf {\bibinfo {volume} {194}},\ \bibinfo {pages} {57}
  (\bibinfo {year} {1982})}\BibitemShut {NoStop}%
\bibitem [{\citenamefont {Semrad}\ \emph {et~al.}(1986)\citenamefont {Semrad},
  \citenamefont {Mertens},\ and\ \citenamefont {Bauer}}]{Semrad:1986}%
  \BibitemOpen
  \bibfield  {author} {\bibinfo {author} {\bibfnamefont {D.}~\bibnamefont
  {Semrad}}, \bibinfo {author} {\bibfnamefont {P.}~\bibnamefont {Mertens}}, \
  and\ \bibinfo {author} {\bibfnamefont {P.}~\bibnamefont {Bauer}},\ }\href
  {\doibase 10.1016/0168-583X(86)90259-4} {\bibfield  {journal} {\bibinfo
  {journal} {Nuclear Instruments and Methods in Physics Research Section B:
  Beam Interactions with Materials and Atoms}\ }\textbf {\bibinfo {volume}
  {15}},\ \bibinfo {pages} {86} (\bibinfo {year} {1986})}\BibitemShut {NoStop}%
\bibitem [{\citenamefont {M{\o}ller}\ \emph {et~al.}(2004)\citenamefont
  {M{\o}ller}, \citenamefont {Csete}, \citenamefont {Ichioka}, \citenamefont
  {Knudsen}, \citenamefont {Uggerh{\o}j},\ and\ \citenamefont
  {Andersen}}]{Moller:2004}%
  \BibitemOpen
  \bibfield  {author} {\bibinfo {author} {\bibfnamefont {S.~P.}\ \bibnamefont
  {M{\o}ller}}, \bibinfo {author} {\bibfnamefont {A.}~\bibnamefont {Csete}},
  \bibinfo {author} {\bibfnamefont {T.}~\bibnamefont {Ichioka}}, \bibinfo
  {author} {\bibfnamefont {H.}~\bibnamefont {Knudsen}}, \bibinfo {author}
  {\bibfnamefont {U.~I.}\ \bibnamefont {Uggerh{\o}j}}, \ and\ \bibinfo {author}
  {\bibfnamefont {H.~H.}\ \bibnamefont {Andersen}},\ }\href {\doibase
  10.1103/PhysRevLett.93.042502} {\bibfield  {journal} {\bibinfo  {journal}
  {Physical Review Letters}\ }\textbf {\bibinfo {volume} {93}},\ \bibinfo
  {pages} {042502} (\bibinfo {year} {2004})}\BibitemShut {NoStop}%
\bibitem [{\citenamefont {Correa}\ \emph {et~al.}(2012)\citenamefont {Correa},
  \citenamefont {Kohanoff}, \citenamefont {Artacho}, \citenamefont
  {Sánchez-Portal},\ and\ \citenamefont {Caro}}]{Correa:2012}%
  \BibitemOpen
  \bibfield  {author} {\bibinfo {author} {\bibfnamefont {A.~A.}\ \bibnamefont
  {Correa}}, \bibinfo {author} {\bibfnamefont {J.}~\bibnamefont {Kohanoff}},
  \bibinfo {author} {\bibfnamefont {E.}~\bibnamefont {Artacho}}, \bibinfo
  {author} {\bibfnamefont {D.}~\bibnamefont {Sánchez-Portal}}, \ and\ \bibinfo
  {author} {\bibfnamefont {A.}~\bibnamefont {Caro}},\ }\href {\doibase
  10.1103/PhysRevLett.108.213201} {\bibfield  {journal} {\bibinfo  {journal}
  {Physical Review Letters}\ }\textbf {\bibinfo {volume} {108}},\ \bibinfo
  {pages} {213201} (\bibinfo {year} {2012})}\BibitemShut {NoStop}%
\bibitem [{\citenamefont {Schleife}\ \emph {et~al.}(2015)\citenamefont
  {Schleife}, \citenamefont {Kanai},\ and\ \citenamefont
  {Correa}}]{Schleife:2015}%
  \BibitemOpen
  \bibfield  {author} {\bibinfo {author} {\bibfnamefont {A.}~\bibnamefont
  {Schleife}}, \bibinfo {author} {\bibfnamefont {Y.}~\bibnamefont {Kanai}}, \
  and\ \bibinfo {author} {\bibfnamefont {A.~A.}\ \bibnamefont {Correa}},\
  }\href {\doibase 10.1103/PhysRevB.91.014306} {\bibfield  {journal} {\bibinfo
  {journal} {Physical Review B}\ }\textbf {\bibinfo {volume} {91}},\ \bibinfo
  {pages} {014306} (\bibinfo {year} {2015})}\BibitemShut {NoStop}%
\bibitem [{\citenamefont {Quashie}\ \emph {et~al.}(2016)\citenamefont
  {Quashie}, \citenamefont {Saha},\ and\ \citenamefont
  {Correa}}]{Quashie:2016}%
  \BibitemOpen
  \bibfield  {author} {\bibinfo {author} {\bibfnamefont {E.~E.}\ \bibnamefont
  {Quashie}}, \bibinfo {author} {\bibfnamefont {B.~C.}\ \bibnamefont {Saha}}, \
  and\ \bibinfo {author} {\bibfnamefont {A.~A.}\ \bibnamefont {Correa}},\
  }\href {\doibase 10.1103/PhysRevB.94.155403} {\bibfield  {journal} {\bibinfo
  {journal} {Physical Review B}\ }\textbf {\bibinfo {volume} {94}},\ \bibinfo
  {pages} {155403} (\bibinfo {year} {2016})}\BibitemShut {NoStop}%
\bibitem [{\citenamefont {Caro}\ \emph {et~al.}(2017)\citenamefont {Caro},
  \citenamefont {Correa}, \citenamefont {Artacho},\ and\ \citenamefont
  {Caro}}]{Caro:2017}%
  \BibitemOpen
  \bibfield  {author} {\bibinfo {author} {\bibfnamefont {M.}~\bibnamefont
  {Caro}}, \bibinfo {author} {\bibfnamefont {A.~A.}\ \bibnamefont {Correa}},
  \bibinfo {author} {\bibfnamefont {E.}~\bibnamefont {Artacho}}, \ and\
  \bibinfo {author} {\bibfnamefont {A.}~\bibnamefont {Caro}},\ }\href {\doibase
  10.1038/s41598-017-02780-3} {\bibfield  {journal} {\bibinfo  {journal}
  {Scientific Reports}\ }\textbf {\bibinfo {volume} {7}},\ \bibinfo {pages}
  {2618} (\bibinfo {year} {2017})}\BibitemShut {NoStop}%
\bibitem [{\citenamefont {Ullah}\ \emph {et~al.}(2018)\citenamefont {Ullah},
  \citenamefont {Artacho},\ and\ \citenamefont {Correa}}]{Ullah:2018}%
  \BibitemOpen
  \bibfield  {author} {\bibinfo {author} {\bibfnamefont {R.}~\bibnamefont
  {Ullah}}, \bibinfo {author} {\bibfnamefont {E.}~\bibnamefont {Artacho}}, \
  and\ \bibinfo {author} {\bibfnamefont {A.~A.}\ \bibnamefont {Correa}},\
  }\href {\doibase 10.1103/PhysRevLett.121.116401} {\bibfield  {journal}
  {\bibinfo  {journal} {Physical Review Letters}\ }\textbf {\bibinfo {volume}
  {121}},\ \bibinfo {pages} {116401} (\bibinfo {year} {2018})}\BibitemShut
  {NoStop}%
\bibitem [{\citenamefont {Lim}\ \emph {et~al.}(2016)\citenamefont {Lim},
  \citenamefont {Foulkes}, \citenamefont {Horsfield}, \citenamefont {Mason},
  \citenamefont {Schleife}, \citenamefont {Draeger},\ and\ \citenamefont
  {Correa}}]{Lim:2016}%
  \BibitemOpen
  \bibfield  {author} {\bibinfo {author} {\bibfnamefont {A.}~\bibnamefont
  {Lim}}, \bibinfo {author} {\bibfnamefont {W.}~\bibnamefont {Foulkes}},
  \bibinfo {author} {\bibfnamefont {A.}~\bibnamefont {Horsfield}}, \bibinfo
  {author} {\bibfnamefont {D.}~\bibnamefont {Mason}}, \bibinfo {author}
  {\bibfnamefont {A.}~\bibnamefont {Schleife}}, \bibinfo {author}
  {\bibfnamefont {E.}~\bibnamefont {Draeger}}, \ and\ \bibinfo {author}
  {\bibfnamefont {A.}~\bibnamefont {Correa}},\ }\href {\doibase
  10.1103/PhysRevLett.116.043201} {\bibfield  {journal} {\bibinfo  {journal}
  {Physical Review Letters}\ }\textbf {\bibinfo {volume} {116}} (\bibinfo
  {year} {2016}),\ 10.1103/PhysRevLett.116.043201}\BibitemShut {NoStop}%
\bibitem [{\citenamefont {Yost}\ and\ \citenamefont {Kanai}(2016)}]{Yost:2016}%
  \BibitemOpen
  \bibfield  {author} {\bibinfo {author} {\bibfnamefont {D.~C.}\ \bibnamefont
  {Yost}}\ and\ \bibinfo {author} {\bibfnamefont {Y.}~\bibnamefont {Kanai}},\
  }\href {\doibase 10.1103/PhysRevB.94.115107} {\bibfield  {journal} {\bibinfo
  {journal} {Physical Review B}\ }\textbf {\bibinfo {volume} {94}} (\bibinfo
  {year} {2016}),\ 10.1103/PhysRevB.94.115107}\BibitemShut {NoStop}%
\bibitem [{\citenamefont {Yost}\ \emph {et~al.}(2017)\citenamefont {Yost},
  \citenamefont {Yao},\ and\ \citenamefont {Kanai}}]{Yost:2017}%
  \BibitemOpen
  \bibfield  {author} {\bibinfo {author} {\bibfnamefont {D.~C.}\ \bibnamefont
  {Yost}}, \bibinfo {author} {\bibfnamefont {Y.}~\bibnamefont {Yao}}, \ and\
  \bibinfo {author} {\bibfnamefont {Y.}~\bibnamefont {Kanai}},\ }\href
  {\doibase 10.1103/PhysRevB.96.115134} {\bibfield  {journal} {\bibinfo
  {journal} {Physical Review B}\ }\textbf {\bibinfo {volume} {96}},\ \bibinfo
  {pages} {115134} (\bibinfo {year} {2017})}\BibitemShut {NoStop}%
\bibitem [{\citenamefont {Lee}\ and\ \citenamefont
  {Schleife}(2018)}]{Lee:2018}%
  \BibitemOpen
  \bibfield  {author} {\bibinfo {author} {\bibfnamefont {C.-W.}\ \bibnamefont
  {Lee}}\ and\ \bibinfo {author} {\bibfnamefont {A.}~\bibnamefont {Schleife}},\
  }\href {\doibase 10.1140/epjb/e2018-90204-8} {\bibfield  {journal} {\bibinfo
  {journal} {The European Physical Journal B}\ }\textbf {\bibinfo {volume}
  {91}},\ \bibinfo {pages} {222} (\bibinfo {year} {2018})}\BibitemShut
  {NoStop}%
\bibitem [{\citenamefont {Kang}\ \emph {et~al.}(2019)\citenamefont {Kang},
  \citenamefont {Kononov}, \citenamefont {Lee}, \citenamefont {Leveillee},
  \citenamefont {Shapera}, \citenamefont {Zhang},\ and\ \citenamefont
  {Schleife}}]{Kang:2019}%
  \BibitemOpen
  \bibfield  {author} {\bibinfo {author} {\bibfnamefont {K.}~\bibnamefont
  {Kang}}, \bibinfo {author} {\bibfnamefont {A.}~\bibnamefont {Kononov}},
  \bibinfo {author} {\bibfnamefont {C.-W.}\ \bibnamefont {Lee}}, \bibinfo
  {author} {\bibfnamefont {J.~A.}\ \bibnamefont {Leveillee}}, \bibinfo {author}
  {\bibfnamefont {E.~P.}\ \bibnamefont {Shapera}}, \bibinfo {author}
  {\bibfnamefont {X.}~\bibnamefont {Zhang}}, \ and\ \bibinfo {author}
  {\bibfnamefont {A.}~\bibnamefont {Schleife}},\ }\href {\doibase
  10.1016/j.commatsci.2019.01.004} {\bibfield  {journal} {\bibinfo  {journal}
  {Computational Materials Science}\ }\textbf {\bibinfo {volume} {160}},\
  \bibinfo {pages} {207} (\bibinfo {year} {2019})}\BibitemShut {NoStop}%
\bibitem [{\citenamefont {Lee}\ and\ \citenamefont
  {Schleife}(2019)}]{Lee:2019}%
  \BibitemOpen
  \bibfield  {author} {\bibinfo {author} {\bibfnamefont {C.-W.}\ \bibnamefont
  {Lee}}\ and\ \bibinfo {author} {\bibfnamefont {A.}~\bibnamefont {Schleife}},\
  }\href {\doibase 10.1021/acs.nanolett.9b01214} {\bibfield  {journal}
  {\bibinfo  {journal} {Nano Letters}\ }\textbf {\bibinfo {volume} {19}},\
  \bibinfo {pages} {3939} (\bibinfo {year} {2019})}\BibitemShut {NoStop}%
\bibitem [{\citenamefont {Pruneda}\ \emph {et~al.}(2007)\citenamefont
  {Pruneda}, \citenamefont {Sánchez-Portal}, \citenamefont {Arnau},
  \citenamefont {Juaristi},\ and\ \citenamefont {Artacho}}]{Pruneda:2007}%
  \BibitemOpen
  \bibfield  {author} {\bibinfo {author} {\bibfnamefont {J.~M.}\ \bibnamefont
  {Pruneda}}, \bibinfo {author} {\bibfnamefont {D.}~\bibnamefont
  {Sánchez-Portal}}, \bibinfo {author} {\bibfnamefont {A.}~\bibnamefont
  {Arnau}}, \bibinfo {author} {\bibfnamefont {J.~I.}\ \bibnamefont {Juaristi}},
  \ and\ \bibinfo {author} {\bibfnamefont {E.}~\bibnamefont {Artacho}},\ }\href
  {\doibase 10.1103/PhysRevLett.99.235501} {\bibfield  {journal} {\bibinfo
  {journal} {Physical Review Letters}\ }\textbf {\bibinfo {volume} {99}},\
  \bibinfo {pages} {235501} (\bibinfo {year} {2007})}\BibitemShut {NoStop}%
\bibitem [{\citenamefont {Mao}\ \emph {et~al.}(2014{\natexlab{a}})\citenamefont
  {Mao}, \citenamefont {Zhang}, \citenamefont {Dai},\ and\ \citenamefont
  {Zhang}}]{Mao:2014:LiF}%
  \BibitemOpen
  \bibfield  {author} {\bibinfo {author} {\bibfnamefont {F.}~\bibnamefont
  {Mao}}, \bibinfo {author} {\bibfnamefont {C.}~\bibnamefont {Zhang}}, \bibinfo
  {author} {\bibfnamefont {J.}~\bibnamefont {Dai}}, \ and\ \bibinfo {author}
  {\bibfnamefont {F.-S.}\ \bibnamefont {Zhang}},\ }\href {\doibase
  10.1103/PhysRevA.89.022707} {\bibfield  {journal} {\bibinfo  {journal}
  {Physical Review A}\ }\textbf {\bibinfo {volume} {89}} (\bibinfo {year}
  {2014}{\natexlab{a}}),\ 10.1103/PhysRevA.89.022707}\BibitemShut {NoStop}%
\bibitem [{\citenamefont {Li}\ \emph {et~al.}(2017)\citenamefont {Li},
  \citenamefont {Wang}, \citenamefont {Liao}, \citenamefont {{OuYang}},\ and\
  \citenamefont {Zhang}}]{Li:2017}%
  \BibitemOpen
  \bibfield  {author} {\bibinfo {author} {\bibfnamefont {C.-K.}\ \bibnamefont
  {Li}}, \bibinfo {author} {\bibfnamefont {F.}~\bibnamefont {Wang}}, \bibinfo
  {author} {\bibfnamefont {B.}~\bibnamefont {Liao}}, \bibinfo {author}
  {\bibfnamefont {X.-P.}\ \bibnamefont {{OuYang}}}, \ and\ \bibinfo {author}
  {\bibfnamefont {F.-S.}\ \bibnamefont {Zhang}},\ }\href {\doibase
  10.1103/PhysRevB.96.094301} {\bibfield  {journal} {\bibinfo  {journal}
  {Physical Review B}\ }\textbf {\bibinfo {volume} {96}},\ \bibinfo {pages}
  {094301} (\bibinfo {year} {2017})}\BibitemShut {NoStop}%
\bibitem [{\citenamefont {Peñalba}\ \emph {et~al.}(1992)\citenamefont
  {Peñalba}, \citenamefont {Arnau}, \citenamefont {Echenique}, \citenamefont
  {Flores},\ and\ \citenamefont {Ritchie}}]{Penalba:1992}%
  \BibitemOpen
  \bibfield  {author} {\bibinfo {author} {\bibfnamefont {M.}~\bibnamefont
  {Peñalba}}, \bibinfo {author} {\bibfnamefont {A.}~\bibnamefont {Arnau}},
  \bibinfo {author} {\bibfnamefont {P.~M.}\ \bibnamefont {Echenique}}, \bibinfo
  {author} {\bibfnamefont {F.}~\bibnamefont {Flores}}, \ and\ \bibinfo {author}
  {\bibfnamefont {R.~H.}\ \bibnamefont {Ritchie}},\ }\href {\doibase
  10.1209/0295-5075/19/1/008} {\bibfield  {journal} {\bibinfo  {journal}
  {Europhysics Letters ({EPL})}\ }\textbf {\bibinfo {volume} {19}},\ \bibinfo
  {pages} {45} (\bibinfo {year} {1992})}\BibitemShut {NoStop}%
\bibitem [{\citenamefont {Quijada}\ \emph {et~al.}(2007)\citenamefont
  {Quijada}, \citenamefont {Borisov}, \citenamefont {Nagy}, \citenamefont
  {Muiño},\ and\ \citenamefont {Echenique}}]{Quijada:2007}%
  \BibitemOpen
  \bibfield  {author} {\bibinfo {author} {\bibfnamefont {M.}~\bibnamefont
  {Quijada}}, \bibinfo {author} {\bibfnamefont {A.~G.}\ \bibnamefont
  {Borisov}}, \bibinfo {author} {\bibfnamefont {I.}~\bibnamefont {Nagy}},
  \bibinfo {author} {\bibfnamefont {R.~D.}\ \bibnamefont {Muiño}}, \ and\
  \bibinfo {author} {\bibfnamefont {P.~M.}\ \bibnamefont {Echenique}},\ }\href
  {\doibase 10.1103/PhysRevA.75.042902} {\bibfield  {journal} {\bibinfo
  {journal} {Physical Review A}\ }\textbf {\bibinfo {volume} {75}},\ \bibinfo
  {pages} {042902} (\bibinfo {year} {2007})}\BibitemShut {NoStop}%
\bibitem [{\citenamefont {Zeb}\ \emph {et~al.}(2013)\citenamefont {Zeb},
  \citenamefont {Kohanoff}, \citenamefont {Sánchez-Portal},\ and\
  \citenamefont {Artacho}}]{Zeb:2013}%
  \BibitemOpen
  \bibfield  {author} {\bibinfo {author} {\bibfnamefont {M.~A.}\ \bibnamefont
  {Zeb}}, \bibinfo {author} {\bibfnamefont {J.}~\bibnamefont {Kohanoff}},
  \bibinfo {author} {\bibfnamefont {D.}~\bibnamefont {Sánchez-Portal}}, \ and\
  \bibinfo {author} {\bibfnamefont {E.}~\bibnamefont {Artacho}},\ }\href
  {\doibase 10.1016/j.nimb.2012.12.022} {\bibfield  {journal} {\bibinfo
  {journal} {Nuclear Instruments and Methods in Physics Research Section B:
  Beam Interactions with Materials and Atoms}\ }\bibinfo {series} {Proceedings
  of the 11th Computer Simulation of Radiation Effects in Solids ({COSIRES})
  Conference Santa Fe, New Mexico, {USA}, July 24-29, 2012},\ \textbf {\bibinfo
  {volume} {303}},\ \bibinfo {pages} {59} (\bibinfo {year} {2013})}\BibitemShut
  {NoStop}%
\bibitem [{\citenamefont {Runge}\ and\ \citenamefont
  {Gross}(1984)}]{Runge:1984}%
  \BibitemOpen
  \bibfield  {author} {\bibinfo {author} {\bibfnamefont {E.}~\bibnamefont
  {Runge}}\ and\ \bibinfo {author} {\bibfnamefont {E.~K.~U.}\ \bibnamefont
  {Gross}},\ }\href {\doibase 10.1103/PhysRevLett.52.997} {\bibfield  {journal}
  {\bibinfo  {journal} {Physical Review Letters}\ }\textbf {\bibinfo {volume}
  {52}},\ \bibinfo {pages} {997} (\bibinfo {year} {1984})}\BibitemShut
  {NoStop}%
\bibitem [{\citenamefont {Marques}\ and\ \citenamefont
  {Gross}(2004)}]{Marques:2004}%
  \BibitemOpen
  \bibfield  {author} {\bibinfo {author} {\bibfnamefont {M.}~\bibnamefont
  {Marques}}\ and\ \bibinfo {author} {\bibfnamefont {E.}~\bibnamefont
  {Gross}},\ }\href {\doibase 10.1146/annurev.physchem.55.091602.094449}
  {\bibfield  {journal} {\bibinfo  {journal} {Annual Review of Physical
  Chemistry}\ }\textbf {\bibinfo {volume} {55}},\ \bibinfo {pages} {427}
  (\bibinfo {year} {2004})}\BibitemShut {NoStop}%
\bibitem [{\citenamefont {Marques}()}]{Marques:2006}%
  \BibitemOpen
  \bibfield  {author} {\bibinfo {author} {\bibfnamefont {M.}~\bibnamefont
  {Marques}},\ }\href@noop {} {\emph {\bibinfo {title} {Time-Dependent Density
  Functional Theory}}}\ (\bibinfo  {publisher} {Springer Science \& Business
  Media})\BibitemShut {NoStop}%
\bibitem [{\citenamefont {Ullrich}(2011)}]{Ullrich:2011}%
  \BibitemOpen
  \bibfield  {author} {\bibinfo {author} {\bibfnamefont {C.~A.}\ \bibnamefont
  {Ullrich}},\ }\href {\doibase 10.1093/acprof:oso/9780199563029.001.0001}
  {\emph {\bibinfo {title} {Time-Dependent Density-Functional Theory: Concepts
  and Applications}}}\ (\bibinfo  {publisher} {Oxford University Press},\
  \bibinfo {year} {2011})\BibitemShut {NoStop}%
\bibitem [{\citenamefont {Ullrich}\ and\ \citenamefont
  {Yang}(2014)}]{Ullrich:2014}%
  \BibitemOpen
  \bibfield  {author} {\bibinfo {author} {\bibfnamefont {C.~A.}\ \bibnamefont
  {Ullrich}}\ and\ \bibinfo {author} {\bibfnamefont {Z.-h.}\ \bibnamefont
  {Yang}},\ }\href {\doibase 10.1007/s13538-013-0141-2} {\bibfield  {journal}
  {\bibinfo  {journal} {Brazilian Journal of Physics}\ }\textbf {\bibinfo
  {volume} {44}},\ \bibinfo {pages} {154} (\bibinfo {year} {2014})}\BibitemShut
  {NoStop}%
\bibitem [{\citenamefont {Schleife}\ \emph {et~al.}(2012)\citenamefont
  {Schleife}, \citenamefont {Draeger}, \citenamefont {Kanai},\ and\
  \citenamefont {Correa}}]{Schleife:2012}%
  \BibitemOpen
  \bibfield  {author} {\bibinfo {author} {\bibfnamefont {A.}~\bibnamefont
  {Schleife}}, \bibinfo {author} {\bibfnamefont {E.~W.}\ \bibnamefont
  {Draeger}}, \bibinfo {author} {\bibfnamefont {Y.}~\bibnamefont {Kanai}}, \
  and\ \bibinfo {author} {\bibfnamefont {A.~A.}\ \bibnamefont {Correa}},\
  }\href {\doibase 10.1063/1.4758792} {\bibfield  {journal} {\bibinfo
  {journal} {The Journal of Chemical Physics}\ }\textbf {\bibinfo {volume}
  {137}},\ \bibinfo {pages} {22A546} (\bibinfo {year} {2012})}\BibitemShut
  {NoStop}%
\bibitem [{\citenamefont {Schleife}\ \emph {et~al.}(2014)\citenamefont
  {Schleife}, \citenamefont {Draeger}, \citenamefont {Anisimov}, \citenamefont
  {Correa},\ and\ \citenamefont {Kanai}}]{Schleife:2014}%
  \BibitemOpen
  \bibfield  {author} {\bibinfo {author} {\bibfnamefont {A.}~\bibnamefont
  {Schleife}}, \bibinfo {author} {\bibfnamefont {E.~W.}\ \bibnamefont
  {Draeger}}, \bibinfo {author} {\bibfnamefont {V.}~\bibnamefont {Anisimov}},
  \bibinfo {author} {\bibfnamefont {A.~A.}\ \bibnamefont {Correa}}, \ and\
  \bibinfo {author} {\bibfnamefont {Y.}~\bibnamefont {Kanai}},\ }\href
  {\doibase 10.1109/MCSE.2014.55} {\bibfield  {journal} {\bibinfo  {journal}
  {Comput. Sci. Eng.}\ }\textbf {\bibinfo {volume} {16}},\ \bibinfo {pages}
  {54} (\bibinfo {year} {2014})},\ \bibinfo {note} {special Issue on
  "Leadership Computing"}\BibitemShut {NoStop}%
\bibitem [{\citenamefont {Gygi}(2008)}]{Gygi:2008}%
  \BibitemOpen
  \bibfield  {author} {\bibinfo {author} {\bibfnamefont {F.}~\bibnamefont
  {Gygi}},\ }\href {\doibase 10.1147/rd.521.0137} {\bibfield  {journal}
  {\bibinfo  {journal} {IBM J. Res. Dev.}\ }\textbf {\bibinfo {volume} {52}},\
  \bibinfo {pages} {137} (\bibinfo {year} {2008})}\BibitemShut {NoStop}%
\bibitem [{\citenamefont {Draeger}\ and\ \citenamefont
  {Gygi}(2017)}]{qball:2017}%
  \BibitemOpen
  \bibfield  {author} {\bibinfo {author} {\bibfnamefont {E.~W.}\ \bibnamefont
  {Draeger}}\ and\ \bibinfo {author} {\bibfnamefont {F.}~\bibnamefont {Gygi}},\
  }\href {https://github.com/LLNL/qball} {\enquote {\bibinfo {title} {Qbox
  code, qb@ll version},}\ } (\bibinfo {year} {2017}),\ \bibinfo {note}
  {{Lawrence Livermore National Laboratory}}\BibitemShut {NoStop}%
\bibitem [{\citenamefont {Troullier}\ and\ \citenamefont
  {Martins}(1991)}]{Troullier:1991}%
  \BibitemOpen
  \bibfield  {author} {\bibinfo {author} {\bibfnamefont {N.}~\bibnamefont
  {Troullier}}\ and\ \bibinfo {author} {\bibfnamefont {J.~L.}\ \bibnamefont
  {Martins}},\ }\href {\doibase 10.1103/PhysRevB.43.1993} {\bibfield  {journal}
  {\bibinfo  {journal} {Physical Review B}\ }\textbf {\bibinfo {volume} {43}},\
  \bibinfo {pages} {1993} (\bibinfo {year} {1991})}\BibitemShut {NoStop}%
\bibitem [{\citenamefont {Vanderbilt}(1985)}]{Vanderbilt:1985}%
  \BibitemOpen
  \bibfield  {author} {\bibinfo {author} {\bibfnamefont {D.}~\bibnamefont
  {Vanderbilt}},\ }\href {\doibase 10.1103/PhysRevB.32.8412} {\bibfield
  {journal} {\bibinfo  {journal} {Physical Review B}\ }\textbf {\bibinfo
  {volume} {32}},\ \bibinfo {pages} {8412} (\bibinfo {year}
  {1985})}\BibitemShut {NoStop}%
\bibitem [{\citenamefont {Zangwill}\ and\ \citenamefont
  {Soven}(1980)}]{Zangwill:1980}%
  \BibitemOpen
  \bibfield  {author} {\bibinfo {author} {\bibfnamefont {A.}~\bibnamefont
  {Zangwill}}\ and\ \bibinfo {author} {\bibfnamefont {P.}~\bibnamefont
  {Soven}},\ }\href {\doibase 10.1103/PhysRevLett.45.204} {\bibfield  {journal}
  {\bibinfo  {journal} {Physical Review Letters}\ }\textbf {\bibinfo {volume}
  {45}},\ \bibinfo {pages} {204} (\bibinfo {year} {1980})}\BibitemShut
  {NoStop}%
\bibitem [{\citenamefont {Zangwill}\ and\ \citenamefont
  {Soven}(1981)}]{Zangwill:1981}%
  \BibitemOpen
  \bibfield  {author} {\bibinfo {author} {\bibfnamefont {A.}~\bibnamefont
  {Zangwill}}\ and\ \bibinfo {author} {\bibfnamefont {P.}~\bibnamefont
  {Soven}},\ }\href {\doibase 10.1103/PhysRevB.24.4121} {\bibfield  {journal}
  {\bibinfo  {journal} {Physical Review B}\ }\textbf {\bibinfo {volume} {24}},\
  \bibinfo {pages} {4121} (\bibinfo {year} {1981})}\BibitemShut {NoStop}%
\bibitem [{\citenamefont {Zhao}\ \emph {et~al.}(2015)\citenamefont {Zhao},
  \citenamefont {Kang}, \citenamefont {Xue}, \citenamefont {Zhang},\ and\
  \citenamefont {Zhang}}]{Zhao:2015}%
  \BibitemOpen
  \bibfield  {author} {\bibinfo {author} {\bibfnamefont {S.}~\bibnamefont
  {Zhao}}, \bibinfo {author} {\bibfnamefont {W.}~\bibnamefont {Kang}}, \bibinfo
  {author} {\bibfnamefont {J.}~\bibnamefont {Xue}}, \bibinfo {author}
  {\bibfnamefont {X.}~\bibnamefont {Zhang}}, \ and\ \bibinfo {author}
  {\bibfnamefont {P.}~\bibnamefont {Zhang}},\ }\href {\doibase
  10.1088/0953-8984/27/2/025401} {\bibfield  {journal} {\bibinfo  {journal}
  {Journal of Physics: Condensed Matter}\ }\textbf {\bibinfo {volume} {27}},\
  \bibinfo {pages} {025401} (\bibinfo {year} {2015})}\BibitemShut {NoStop}%
\bibitem [{\citenamefont {Castro}\ \emph {et~al.}(2004)\citenamefont {Castro},
  \citenamefont {Marques},\ and\ \citenamefont {Rubio}}]{Castro:2004}%
  \BibitemOpen
  \bibfield  {author} {\bibinfo {author} {\bibfnamefont {A.}~\bibnamefont
  {Castro}}, \bibinfo {author} {\bibfnamefont {M.~A.~L.}\ \bibnamefont
  {Marques}}, \ and\ \bibinfo {author} {\bibfnamefont {A.}~\bibnamefont
  {Rubio}},\ }\href {\doibase 10.1063/1.1774980} {\bibfield  {journal}
  {\bibinfo  {journal} {The Journal of Chemical Physics}\ }\textbf {\bibinfo
  {volume} {121}},\ \bibinfo {pages} {3425} (\bibinfo {year}
  {2004})}\BibitemShut {NoStop}%
\bibitem [{\citenamefont {Draeger}\ \emph {et~al.}(2017)\citenamefont
  {Draeger}, \citenamefont {Andrade}, \citenamefont {Gunnels}, \citenamefont
  {Bhatele}, \citenamefont {Schleife},\ and\ \citenamefont
  {Correa}}]{Draeger:2017}%
  \BibitemOpen
  \bibfield  {author} {\bibinfo {author} {\bibfnamefont {E.~W.}\ \bibnamefont
  {Draeger}}, \bibinfo {author} {\bibfnamefont {X.}~\bibnamefont {Andrade}},
  \bibinfo {author} {\bibfnamefont {J.~A.}\ \bibnamefont {Gunnels}}, \bibinfo
  {author} {\bibfnamefont {A.}~\bibnamefont {Bhatele}}, \bibinfo {author}
  {\bibfnamefont {A.}~\bibnamefont {Schleife}}, \ and\ \bibinfo {author}
  {\bibfnamefont {A.~A.}\ \bibnamefont {Correa}},\ }\href {\doibase
  10.1016/j.jpdc.2017.02.005} {\bibfield  {journal} {\bibinfo  {journal}
  {Journal of Parallel and Distributed Computing}\ }\textbf {\bibinfo {volume}
  {106}},\ \bibinfo {pages} {205} (\bibinfo {year} {2017})}\BibitemShut
  {NoStop}%
\bibitem [{\citenamefont {Blaiszik}\ \emph {et~al.}(2016)\citenamefont
  {Blaiszik}, \citenamefont {Chard}, \citenamefont {Pruyne}, \citenamefont
  {Ananthakrishnan}, \citenamefont {Tuecke},\ and\ \citenamefont
  {Foster}}]{MDF}%
  \BibitemOpen
  \bibfield  {author} {\bibinfo {author} {\bibfnamefont {B.}~\bibnamefont
  {Blaiszik}}, \bibinfo {author} {\bibfnamefont {K.}~\bibnamefont {Chard}},
  \bibinfo {author} {\bibfnamefont {J.}~\bibnamefont {Pruyne}}, \bibinfo
  {author} {\bibfnamefont {R.}~\bibnamefont {Ananthakrishnan}}, \bibinfo
  {author} {\bibfnamefont {S.}~\bibnamefont {Tuecke}}, \ and\ \bibinfo {author}
  {\bibfnamefont {I.}~\bibnamefont {Foster}},\ }\href {\doibase
  10.1007/s11837-016-2001-3} {\bibfield  {journal} {\bibinfo  {journal} {JOM}\
  }\textbf {\bibinfo {volume} {68}},\ \bibinfo {pages} {2045} (\bibinfo {year}
  {2016})}\BibitemShut {NoStop}%
\bibitem [{\citenamefont {Kononov}\ and\ \citenamefont
  {Schleife}(2020)}]{data}%
  \BibitemOpen
  \bibfield  {author} {\bibinfo {author} {\bibfnamefont {A.}~\bibnamefont
  {Kononov}}\ and\ \bibinfo {author} {\bibfnamefont {A.}~\bibnamefont
  {Schleife}},\ }\href {\doibase 10.18126/4ene-mwpd} {\enquote {\bibinfo
  {title} {Dataset for `pre-equilibrium stopping and charge capture in
  proton-irradiated aluminum sheets'},}\ }\bibinfo {howpublished}
  {\url{https://www.doi.org/10.18126/4ene-mwpd}} (\bibinfo {year}
  {2020})\BibitemShut {NoStop}%
\bibitem [{\citenamefont {Mulliken}(1955)}]{Mulliken:1955}%
  \BibitemOpen
  \bibfield  {author} {\bibinfo {author} {\bibfnamefont {R.~S.}\ \bibnamefont
  {Mulliken}},\ }\href {\doibase 10.1063/1.1740588} {\bibfield  {journal}
  {\bibinfo  {journal} {The Journal of Chemical Physics}\ }\textbf {\bibinfo
  {volume} {23}},\ \bibinfo {pages} {1833} (\bibinfo {year}
  {1955})}\BibitemShut {NoStop}%
\bibitem [{\citenamefont {Knospe}\ \emph {et~al.}(1999)\citenamefont {Knospe},
  \citenamefont {Jellinek}, \citenamefont {Saalmann},\ and\ \citenamefont
  {Schmidt}}]{Knospe:1999}%
  \BibitemOpen
  \bibfield  {author} {\bibinfo {author} {\bibfnamefont {O.}~\bibnamefont
  {Knospe}}, \bibinfo {author} {\bibfnamefont {J.}~\bibnamefont {Jellinek}},
  \bibinfo {author} {\bibfnamefont {U.}~\bibnamefont {Saalmann}}, \ and\
  \bibinfo {author} {\bibfnamefont {R.}~\bibnamefont {Schmidt}},\ }\href
  {\doibase 10.1007/PL00021588} {\bibfield  {journal} {\bibinfo  {journal} {The
  European Physical Journal D - Atomic, Molecular, Optical and Plasma Physics}\
  }\textbf {\bibinfo {volume} {5}},\ \bibinfo {pages} {1} (\bibinfo {year}
  {1999})}\BibitemShut {NoStop}%
\bibitem [{\citenamefont {Knospe}\ \emph {et~al.}(2000)\citenamefont {Knospe},
  \citenamefont {Jellinek}, \citenamefont {Saalmann},\ and\ \citenamefont
  {Schmidt}}]{Knospe:2000}%
  \BibitemOpen
  \bibfield  {author} {\bibinfo {author} {\bibfnamefont {O.}~\bibnamefont
  {Knospe}}, \bibinfo {author} {\bibfnamefont {J.}~\bibnamefont {Jellinek}},
  \bibinfo {author} {\bibfnamefont {U.}~\bibnamefont {Saalmann}}, \ and\
  \bibinfo {author} {\bibfnamefont {R.}~\bibnamefont {Schmidt}},\ }\href
  {\doibase 10.1103/PhysRevA.61.022715} {\bibfield  {journal} {\bibinfo
  {journal} {Physical Review A}\ }\textbf {\bibinfo {volume} {61}},\ \bibinfo
  {pages} {022715} (\bibinfo {year} {2000})}\BibitemShut {NoStop}%
\bibitem [{\citenamefont {Isborn}\ \emph {et~al.}(2007)\citenamefont {Isborn},
  \citenamefont {Li},\ and\ \citenamefont {Tully}}]{Isborn:2007}%
  \BibitemOpen
  \bibfield  {author} {\bibinfo {author} {\bibfnamefont {C.~M.}\ \bibnamefont
  {Isborn}}, \bibinfo {author} {\bibfnamefont {X.}~\bibnamefont {Li}}, \ and\
  \bibinfo {author} {\bibfnamefont {J.~C.}\ \bibnamefont {Tully}},\ }\href
  {\doibase 10.1063/1.2713391} {\bibfield  {journal} {\bibinfo  {journal} {The
  Journal of Chemical Physics}\ }\textbf {\bibinfo {volume} {126}},\ \bibinfo
  {pages} {134307} (\bibinfo {year} {2007})}\BibitemShut {NoStop}%
\bibitem [{\citenamefont {Miyamoto}\ and\ \citenamefont
  {Zhang}(2008{\natexlab{a}})}]{Miyamoto:2008}%
  \BibitemOpen
  \bibfield  {author} {\bibinfo {author} {\bibfnamefont {Y.}~\bibnamefont
  {Miyamoto}}\ and\ \bibinfo {author} {\bibfnamefont {H.}~\bibnamefont
  {Zhang}},\ }\href {\doibase 10.1103/PhysRevB.77.161402} {\bibfield  {journal}
  {\bibinfo  {journal} {Physical Review B}\ }\textbf {\bibinfo {volume} {77}},\
  \bibinfo {pages} {161402} (\bibinfo {year} {2008}{\natexlab{a}})}\BibitemShut
  {NoStop}%
\bibitem [{\citenamefont {Reinhard}\ and\ \citenamefont
  {Suraud}(1999)}]{Reinhard:1999}%
  \BibitemOpen
  \bibfield  {author} {\bibinfo {author} {\bibfnamefont {P.-G.}\ \bibnamefont
  {Reinhard}}\ and\ \bibinfo {author} {\bibfnamefont {E.}~\bibnamefont
  {Suraud}},\ }\href {\doibase 10.1023/A:1021973512410} {\bibfield  {journal}
  {\bibinfo  {journal} {Journal of Cluster Science}\ }\textbf {\bibinfo
  {volume} {10}},\ \bibinfo {pages} {239} (\bibinfo {year} {1999})}\BibitemShut
  {NoStop}%
\bibitem [{\citenamefont {Ullrich}(2000)}]{Ullrich:2000}%
  \BibitemOpen
  \bibfield  {author} {\bibinfo {author} {\bibfnamefont {C.~A.}\ \bibnamefont
  {Ullrich}},\ }\href {\doibase 10.1016/S0166-1280(99)00442-X} {\bibfield
  {journal} {\bibinfo  {journal} {Journal of Molecular Structure: {THEOCHEM}}\
  }\textbf {\bibinfo {volume} {501-502}},\ \bibinfo {pages} {315} (\bibinfo
  {year} {2000})}\BibitemShut {NoStop}%
\bibitem [{\citenamefont {Wang}\ \emph {et~al.}(2011)\citenamefont {Wang},
  \citenamefont {Xu}, \citenamefont {Hong}, \citenamefont {Wang},\ and\
  \citenamefont {Gou}}]{Wang:2011}%
  \BibitemOpen
  \bibfield  {author} {\bibinfo {author} {\bibfnamefont {F.}~\bibnamefont
  {Wang}}, \bibinfo {author} {\bibfnamefont {X.~C.}\ \bibnamefont {Xu}},
  \bibinfo {author} {\bibfnamefont {X.~H.}\ \bibnamefont {Hong}}, \bibinfo
  {author} {\bibfnamefont {J.}~\bibnamefont {Wang}}, \ and\ \bibinfo {author}
  {\bibfnamefont {B.~C.}\ \bibnamefont {Gou}},\ }\href {\doibase
  10.1016/j.physleta.2011.07.032} {\bibfield  {journal} {\bibinfo  {journal}
  {Physics Letters A}\ }\textbf {\bibinfo {volume} {375}},\ \bibinfo {pages}
  {3290} (\bibinfo {year} {2011})}\BibitemShut {NoStop}%
\bibitem [{\citenamefont {Wang}\ \emph {et~al.}(2012)\citenamefont {Wang},
  \citenamefont {Hong}, \citenamefont {Wang}, \citenamefont {Gou},\ and\
  \citenamefont {Wang}}]{Wang:2012}%
  \BibitemOpen
  \bibfield  {author} {\bibinfo {author} {\bibfnamefont {F.}~\bibnamefont
  {Wang}}, \bibinfo {author} {\bibfnamefont {X.~H.}\ \bibnamefont {Hong}},
  \bibinfo {author} {\bibfnamefont {J.}~\bibnamefont {Wang}}, \bibinfo {author}
  {\bibfnamefont {B.~C.}\ \bibnamefont {Gou}}, \ and\ \bibinfo {author}
  {\bibfnamefont {J.~G.}\ \bibnamefont {Wang}},\ }\href {\doibase
  10.1016/j.physleta.2011.11.031} {\bibfield  {journal} {\bibinfo  {journal}
  {Physics Letters A}\ }\textbf {\bibinfo {volume} {376}},\ \bibinfo {pages}
  {469} (\bibinfo {year} {2012})}\BibitemShut {NoStop}%
\bibitem [{\citenamefont {Gao}\ \emph {et~al.}(2013)\citenamefont {Gao},
  \citenamefont {Wang},\ and\ \citenamefont {Zhang}}]{Gao:2013}%
  \BibitemOpen
  \bibfield  {author} {\bibinfo {author} {\bibfnamefont {C.-Z.}\ \bibnamefont
  {Gao}}, \bibinfo {author} {\bibfnamefont {J.}~\bibnamefont {Wang}}, \ and\
  \bibinfo {author} {\bibfnamefont {F.-S.}\ \bibnamefont {Zhang}},\ }\href
  {\doibase 10.1016/j.chemphys.2012.10.007} {\bibfield  {journal} {\bibinfo
  {journal} {Chemical Physics}\ }\textbf {\bibinfo {volume} {410}},\ \bibinfo
  {pages} {9} (\bibinfo {year} {2013})}\BibitemShut {NoStop}%
\bibitem [{\citenamefont {Bader}(1994)}]{Bader:1994}%
  \BibitemOpen
  \bibfield  {author} {\bibinfo {author} {\bibfnamefont {R.~F.~W.}\
  \bibnamefont {Bader}},\ }\href@noop {} {\emph {\bibinfo {title} {Atoms in
  Molecules: A Quantum Theory}}}\ (\bibinfo  {publisher} {Clarendon Press},\
  \bibinfo {year} {1994})\BibitemShut {NoStop}%
\bibitem [{\citenamefont {Manz}\ and\ \citenamefont {Limas}(2016)}]{Manz:2016}%
  \BibitemOpen
  \bibfield  {author} {\bibinfo {author} {\bibfnamefont {T.~A.}\ \bibnamefont
  {Manz}}\ and\ \bibinfo {author} {\bibfnamefont {N.~G.}\ \bibnamefont
  {Limas}},\ }\href {\doibase 10.1039/C6RA04656H} {\bibfield  {journal}
  {\bibinfo  {journal} {{RSC} Advances}\ }\textbf {\bibinfo {volume} {6}},\
  \bibinfo {pages} {47771} (\bibinfo {year} {2016})}\BibitemShut {NoStop}%
\bibitem [{\citenamefont {Gabaldon~Limas}\ and\ \citenamefont
  {Manz}(2016)}]{Gabaldon_limas:2016}%
  \BibitemOpen
  \bibfield  {author} {\bibinfo {author} {\bibfnamefont {N.}~\bibnamefont
  {Gabaldon~Limas}}\ and\ \bibinfo {author} {\bibfnamefont {T.~A.}\
  \bibnamefont {Manz}},\ }\href {\doibase 10.1039/C6RA05507A} {\bibfield
  {journal} {\bibinfo  {journal} {{RSC} Advances}\ }\textbf {\bibinfo {volume}
  {6}},\ \bibinfo {pages} {45727} (\bibinfo {year} {2016})}\BibitemShut
  {NoStop}%
\bibitem [{\citenamefont {Reinhard}\ \emph {et~al.}(1998)\citenamefont
  {Reinhard}, \citenamefont {Suraud},\ and\ \citenamefont
  {Ullrich}}]{Reinhard:1998}%
  \BibitemOpen
  \bibfield  {author} {\bibinfo {author} {\bibfnamefont {P.}~\bibnamefont
  {Reinhard}}, \bibinfo {author} {\bibfnamefont {E.}~\bibnamefont {Suraud}}, \
  and\ \bibinfo {author} {\bibfnamefont {C.}~\bibnamefont {Ullrich}},\ }\href
  {\doibase 10.1007/s100530050097} {\bibfield  {journal} {\bibinfo  {journal}
  {The European Physical Journal D - Atomic, Molecular, Optical and Plasma
  Physics}\ }\textbf {\bibinfo {volume} {1}},\ \bibinfo {pages} {303} (\bibinfo
  {year} {1998})}\BibitemShut {NoStop}%
\bibitem [{\citenamefont {Miyamoto}\ and\ \citenamefont
  {Zhang}(2008{\natexlab{b}})}]{Miyamoto:2008:general}%
  \BibitemOpen
  \bibfield  {author} {\bibinfo {author} {\bibfnamefont {Y.}~\bibnamefont
  {Miyamoto}}\ and\ \bibinfo {author} {\bibfnamefont {H.}~\bibnamefont
  {Zhang}},\ }\href {\doibase 10.1103/PhysRevB.77.045433} {\bibfield  {journal}
  {\bibinfo  {journal} {Physical Review B}\ }\textbf {\bibinfo {volume} {77}},\
  \bibinfo {pages} {045433} (\bibinfo {year} {2008}{\natexlab{b}})}\BibitemShut
  {NoStop}%
\bibitem [{\citenamefont {Avendano-Franco}\ \emph {et~al.}(2012)\citenamefont
  {Avendano-Franco}, \citenamefont {Piraux}, \citenamefont {Grüning},\ and\
  \citenamefont {Gonze}}]{Avendano:2012}%
  \BibitemOpen
  \bibfield  {author} {\bibinfo {author} {\bibfnamefont {G.}~\bibnamefont
  {Avendano-Franco}}, \bibinfo {author} {\bibfnamefont {B.}~\bibnamefont
  {Piraux}}, \bibinfo {author} {\bibfnamefont {M.}~\bibnamefont {Grüning}}, \
  and\ \bibinfo {author} {\bibfnamefont {X.}~\bibnamefont {Gonze}},\ }\href
  {\doibase 10.1007/s00214-012-1289-5} {\bibfield  {journal} {\bibinfo
  {journal} {Theoretical Chemistry Accounts}\ }\textbf {\bibinfo {volume}
  {131}},\ \bibinfo {pages} {1289} (\bibinfo {year} {2012})}\BibitemShut
  {NoStop}%
\bibitem [{\citenamefont {Mao}\ \emph {et~al.}(2014{\natexlab{b}})\citenamefont
  {Mao}, \citenamefont {Zhang}, \citenamefont {Gao}, \citenamefont {Dai},\ and\
  \citenamefont {Zhang}}]{Mao:2014:Gr}%
  \BibitemOpen
  \bibfield  {author} {\bibinfo {author} {\bibfnamefont {F.}~\bibnamefont
  {Mao}}, \bibinfo {author} {\bibfnamefont {C.}~\bibnamefont {Zhang}}, \bibinfo
  {author} {\bibfnamefont {C.-Z.}\ \bibnamefont {Gao}}, \bibinfo {author}
  {\bibfnamefont {J.}~\bibnamefont {Dai}}, \ and\ \bibinfo {author}
  {\bibfnamefont {F.-S.}\ \bibnamefont {Zhang}},\ }\href {\doibase
  10.1088/0953-8984/26/8/085402} {\bibfield  {journal} {\bibinfo  {journal}
  {Journal of Physics: Condensed Matter}\ }\textbf {\bibinfo {volume} {26}},\
  \bibinfo {pages} {085402} (\bibinfo {year} {2014}{\natexlab{b}})}\BibitemShut
  {NoStop}%
\bibitem [{\citenamefont {Ojanperä}\ \emph {et~al.}(2014)\citenamefont
  {Ojanperä}, \citenamefont {Krasheninnikov},\ and\ \citenamefont
  {Puska}}]{Ojanpera:2014}%
  \BibitemOpen
  \bibfield  {author} {\bibinfo {author} {\bibfnamefont {A.}~\bibnamefont
  {Ojanperä}}, \bibinfo {author} {\bibfnamefont {A.~V.}\ \bibnamefont
  {Krasheninnikov}}, \ and\ \bibinfo {author} {\bibfnamefont {M.}~\bibnamefont
  {Puska}},\ }\href {\doibase 10.1103/PhysRevB.89.035120} {\bibfield  {journal}
  {\bibinfo  {journal} {Physical Review B}\ }\textbf {\bibinfo {volume} {89}},\
  \bibinfo {pages} {035120} (\bibinfo {year} {2014})}\BibitemShut {NoStop}%
\bibitem [{\citenamefont {Herrmann}\ \emph {et~al.}(1994)\citenamefont
  {Herrmann}, \citenamefont {Cocke}, \citenamefont {Ullrich}, \citenamefont
  {Hagmann}, \citenamefont {Stoeckli},\ and\ \citenamefont
  {Schmidt-Boecking}}]{Herrmann:1994}%
  \BibitemOpen
  \bibfield  {author} {\bibinfo {author} {\bibfnamefont {R.}~\bibnamefont
  {Herrmann}}, \bibinfo {author} {\bibfnamefont {C.~L.}\ \bibnamefont {Cocke}},
  \bibinfo {author} {\bibfnamefont {J.}~\bibnamefont {Ullrich}}, \bibinfo
  {author} {\bibfnamefont {S.}~\bibnamefont {Hagmann}}, \bibinfo {author}
  {\bibfnamefont {M.}~\bibnamefont {Stoeckli}}, \ and\ \bibinfo {author}
  {\bibfnamefont {H.}~\bibnamefont {Schmidt-Boecking}},\ }\href {\doibase
  10.1103/PhysRevA.50.1435} {\bibfield  {journal} {\bibinfo  {journal}
  {Physical Review A}\ }\textbf {\bibinfo {volume} {50}},\ \bibinfo {pages}
  {1435} (\bibinfo {year} {1994})}\BibitemShut {NoStop}%
\bibitem [{\citenamefont {Bohr}(1948)}]{Bohr:1948}%
  \BibitemOpen
  \bibfield  {author} {\bibinfo {author} {\bibfnamefont {N.}~\bibnamefont
  {Bohr}},\ }\href@noop {} {\emph {\bibinfo {title} {The Penetration Atomic
  Particles Through Matter}}}\ (\bibinfo  {publisher} {The Royal Danish Academy
  of Science},\ \bibinfo {year} {1948})\BibitemShut {NoStop}%
\bibitem [{\citenamefont {Ogawa}\ \emph {et~al.}(1993)\citenamefont {Ogawa},
  \citenamefont {Katayama}, \citenamefont {Sugai}, \citenamefont {Haruyama},
  \citenamefont {Saito}, \citenamefont {Yoshida}, \citenamefont {Tosaki},\ and\
  \citenamefont {Ikegami}}]{Ogawa:1993}%
  \BibitemOpen
  \bibfield  {author} {\bibinfo {author} {\bibfnamefont {H.}~\bibnamefont
  {Ogawa}}, \bibinfo {author} {\bibfnamefont {I.}~\bibnamefont {Katayama}},
  \bibinfo {author} {\bibfnamefont {I.}~\bibnamefont {Sugai}}, \bibinfo
  {author} {\bibfnamefont {Y.}~\bibnamefont {Haruyama}}, \bibinfo {author}
  {\bibfnamefont {M.}~\bibnamefont {Saito}}, \bibinfo {author} {\bibfnamefont
  {K.}~\bibnamefont {Yoshida}}, \bibinfo {author} {\bibfnamefont
  {M.}~\bibnamefont {Tosaki}}, \ and\ \bibinfo {author} {\bibfnamefont
  {H.}~\bibnamefont {Ikegami}},\ }\href {\doibase 10.1016/0168-583X(93)95085-J}
  {\bibfield  {journal} {\bibinfo  {journal} {Nuclear Instruments and Methods
  in Physics Research Section B: Beam Interactions with Materials and Atoms}\
  }\textbf {\bibinfo {volume} {82}},\ \bibinfo {pages} {80} (\bibinfo {year}
  {1993})}\BibitemShut {NoStop}%
\bibitem [{\citenamefont {Ziegler}\ \emph {et~al.}(2010)\citenamefont
  {Ziegler}, \citenamefont {Ziegler},\ and\ \citenamefont
  {Biersack}}]{Ziegler:2010}%
  \BibitemOpen
  \bibfield  {author} {\bibinfo {author} {\bibfnamefont {J.~F.}\ \bibnamefont
  {Ziegler}}, \bibinfo {author} {\bibfnamefont {M.~D.}\ \bibnamefont
  {Ziegler}}, \ and\ \bibinfo {author} {\bibfnamefont {J.~P.}\ \bibnamefont
  {Biersack}},\ }\href {\doibase 10.1016/j.nimb.2010.02.091} {\bibfield
  {journal} {\bibinfo  {journal} {Nuclear Instruments and Methods in Physics
  Research Section B: Beam Interactions with Materials and Atoms}\ }\bibinfo
  {series} {19th International Conference on Ion Beam Analysis},\ \textbf
  {\bibinfo {volume} {268}},\ \bibinfo {pages} {1818} (\bibinfo {year}
  {2010})}\BibitemShut {NoStop}%
\bibitem [{\citenamefont {Pines}(1956)}]{Pines:1956}%
  \BibitemOpen
  \bibfield  {author} {\bibinfo {author} {\bibfnamefont {D.}~\bibnamefont
  {Pines}},\ }\href {\doibase 10.1103/RevModPhys.28.184} {\bibfield  {journal}
  {\bibinfo  {journal} {Reviews of Modern Physics}\ }\textbf {\bibinfo {volume}
  {28}},\ \bibinfo {pages} {184} (\bibinfo {year} {1956})}\BibitemShut
  {NoStop}%
\bibitem [{\citenamefont {Powell}\ and\ \citenamefont
  {Swan}(1959)}]{Powell:1959}%
  \BibitemOpen
  \bibfield  {author} {\bibinfo {author} {\bibfnamefont {C.~J.}\ \bibnamefont
  {Powell}}\ and\ \bibinfo {author} {\bibfnamefont {J.~B.}\ \bibnamefont
  {Swan}},\ }\href {\doibase 10.1103/PhysRev.115.869} {\bibfield  {journal}
  {\bibinfo  {journal} {Physical Review}\ }\textbf {\bibinfo {volume} {115}},\
  \bibinfo {pages} {869} (\bibinfo {year} {1959})}\BibitemShut {NoStop}%
\bibitem [{\citenamefont {Grosso}\ and\ \citenamefont
  {Parravicini}(2014)}]{Grosso:2014}%
  \BibitemOpen
  \bibfield  {author} {\bibinfo {author} {\bibfnamefont {G.}~\bibnamefont
  {Grosso}}\ and\ \bibinfo {author} {\bibfnamefont {G.~P.}\ \bibnamefont
  {Parravicini}},\ }in\ \href {\doibase 10.1016/B978-0-12-385030-0.00007-4}
  {\emph {\bibinfo {booktitle} {Solid State Physics (Second Edition)}}}\
  (\bibinfo  {publisher} {Academic Press},\ \bibinfo {year} {2014})\ pp.\
  \bibinfo {pages} {287--331}\BibitemShut {NoStop}%
\bibitem [{\citenamefont {Maniyara}\ \emph {et~al.}(2019)\citenamefont
  {Maniyara}, \citenamefont {Rodrigo}, \citenamefont {Yu}, \citenamefont
  {Canet-Ferrer}, \citenamefont {Ghosh}, \citenamefont {Yongsunthon},
  \citenamefont {Baker}, \citenamefont {Rezikyan}, \citenamefont {García~de
  Abajo},\ and\ \citenamefont {Pruneri}}]{Maniyara:2019}%
  \BibitemOpen
  \bibfield  {author} {\bibinfo {author} {\bibfnamefont {R.~A.}\ \bibnamefont
  {Maniyara}}, \bibinfo {author} {\bibfnamefont {D.}~\bibnamefont {Rodrigo}},
  \bibinfo {author} {\bibfnamefont {R.}~\bibnamefont {Yu}}, \bibinfo {author}
  {\bibfnamefont {J.}~\bibnamefont {Canet-Ferrer}}, \bibinfo {author}
  {\bibfnamefont {D.~S.}\ \bibnamefont {Ghosh}}, \bibinfo {author}
  {\bibfnamefont {R.}~\bibnamefont {Yongsunthon}}, \bibinfo {author}
  {\bibfnamefont {D.~E.}\ \bibnamefont {Baker}}, \bibinfo {author}
  {\bibfnamefont {A.}~\bibnamefont {Rezikyan}}, \bibinfo {author}
  {\bibfnamefont {F.~J.}\ \bibnamefont {García~de Abajo}}, \ and\ \bibinfo
  {author} {\bibfnamefont {V.}~\bibnamefont {Pruneri}},\ }\href {\doibase
  10.1038/s41566-019-0366-x} {\bibfield  {journal} {\bibinfo  {journal} {Nature
  Photonics}\ }\textbf {\bibinfo {volume} {13}},\ \bibinfo {pages} {328}
  (\bibinfo {year} {2019})}\BibitemShut {NoStop}%
\bibitem [{\citenamefont {Andersen}\ \emph {et~al.}(2012)\citenamefont
  {Andersen}, \citenamefont {Jacobsen},\ and\ \citenamefont
  {Thygesen}}]{Andersen:2012}%
  \BibitemOpen
  \bibfield  {author} {\bibinfo {author} {\bibfnamefont {K.}~\bibnamefont
  {Andersen}}, \bibinfo {author} {\bibfnamefont {K.~W.}\ \bibnamefont
  {Jacobsen}}, \ and\ \bibinfo {author} {\bibfnamefont {K.~S.}\ \bibnamefont
  {Thygesen}},\ }\href {\doibase 10.1103/PhysRevB.86.245129} {\bibfield
  {journal} {\bibinfo  {journal} {Physical Review B}\ }\textbf {\bibinfo
  {volume} {86}},\ \bibinfo {pages} {245129} (\bibinfo {year}
  {2012})}\BibitemShut {NoStop}%
\bibitem [{\citenamefont {Ritchie}(1957)}]{Ritchie:1957}%
  \BibitemOpen
  \bibfield  {author} {\bibinfo {author} {\bibfnamefont {R.~H.}\ \bibnamefont
  {Ritchie}},\ }\href {\doibase 10.1103/PhysRev.106.874} {\bibfield  {journal}
  {\bibinfo  {journal} {Physical Review}\ }\textbf {\bibinfo {volume} {106}},\
  \bibinfo {pages} {874} (\bibinfo {year} {1957})}\BibitemShut {NoStop}%
\bibitem [{\citenamefont {Bürgi}\ \emph {et~al.}(1993)\citenamefont {Bürgi},
  \citenamefont {Gonin}, \citenamefont {Oetliker}, \citenamefont {Bochsler},
  \citenamefont {Geiss}, \citenamefont {Lamy}, \citenamefont {Brenac},
  \citenamefont {Andrä}, \citenamefont {Roncin}, \citenamefont {Laurent},\
  and\ \citenamefont {Coplan}}]{Burgi:1993}%
  \BibitemOpen
  \bibfield  {author} {\bibinfo {author} {\bibfnamefont {A.}~\bibnamefont
  {Bürgi}}, \bibinfo {author} {\bibfnamefont {M.}~\bibnamefont {Gonin}},
  \bibinfo {author} {\bibfnamefont {M.}~\bibnamefont {Oetliker}}, \bibinfo
  {author} {\bibfnamefont {P.}~\bibnamefont {Bochsler}}, \bibinfo {author}
  {\bibfnamefont {J.}~\bibnamefont {Geiss}}, \bibinfo {author} {\bibfnamefont
  {T.}~\bibnamefont {Lamy}}, \bibinfo {author} {\bibfnamefont {A.}~\bibnamefont
  {Brenac}}, \bibinfo {author} {\bibfnamefont {H.~J.}\ \bibnamefont {Andrä}},
  \bibinfo {author} {\bibfnamefont {P.}~\bibnamefont {Roncin}}, \bibinfo
  {author} {\bibfnamefont {H.}~\bibnamefont {Laurent}}, \ and\ \bibinfo
  {author} {\bibfnamefont {M.~A.}\ \bibnamefont {Coplan}},\ }\href {\doibase
  10.1063/1.352846} {\bibfield  {journal} {\bibinfo  {journal} {Journal of
  Applied Physics}\ }\textbf {\bibinfo {volume} {73}},\ \bibinfo {pages} {4130}
  (\bibinfo {year} {1993})}\BibitemShut {NoStop}%
\bibitem [{\citenamefont {Folkerts}\ \emph {et~al.}(1995)\citenamefont
  {Folkerts}, \citenamefont {Schippers}, \citenamefont {Zehner},\ and\
  \citenamefont {Meyer}}]{Folkerts:1995}%
  \BibitemOpen
  \bibfield  {author} {\bibinfo {author} {\bibfnamefont {L.}~\bibnamefont
  {Folkerts}}, \bibinfo {author} {\bibfnamefont {S.}~\bibnamefont {Schippers}},
  \bibinfo {author} {\bibfnamefont {D.~M.}\ \bibnamefont {Zehner}}, \ and\
  \bibinfo {author} {\bibfnamefont {F.~W.}\ \bibnamefont {Meyer}},\ }\href
  {\doibase 10.1103/PhysRevLett.74.2204} {\bibfield  {journal} {\bibinfo
  {journal} {Physical Review Letters}\ }\textbf {\bibinfo {volume} {74}},\
  \bibinfo {pages} {2204} (\bibinfo {year} {1995})}\BibitemShut {NoStop}%
\end{thebibliography}%
\end{document}